﻿

\documentclass[traditabstract,a4paper]{aa}
\usepackage{url}
\usepackage{times}
\usepackage{graphicx}
\newcommand{\MSun}{\mbox{M$_\odot$}}

\def\apgt{\ {\raise-.5ex\hbox{$\buildrel>\over\sim$}}\ }
\def\aplt{\ {\raise-.5ex\hbox{$\buildrel<\over\sim$}}\ }
\def\lteq{\ {\raise-.5ex\hbox{$\buildrel<\over-$}}\ }

\usepackage{color}

\newcommand{\Solus}{S\={o}lus lapis}
\newcommand{\solus}{s\={o}lus lapis}
\newcommand{\soli}{s\={o}l\={\i} lapid\={e}s}
\newcommand{\Soli}{s\={o}l\={\i} lapid\={e}s}

\begin{document}
\title{Oort cloud Ecology II:
       Extra-solar Oort clouds and the origin of asteroidal interlopers}
\author{S.\, Portegies Zwart\inst{1}}


\offprints{S. Portegies Zwart}
\mail{spz@strw.leidenuniv.nl}
\institute{
$^1$Leiden Observatory, Leiden University, PO Box 9513, 2300
RA, Leiden, The Netherlands
}
\date{Received / Accepted }
\titlerunning{Oort cloud Ecology II}
\authorrunning{Portegies Zwart et al.}

\abstract{ We simulate the formation and evolution of Oort clouds
  around the 200 nearest stars (within $\sim 16$\,pc according to the
  Gaia DR2) database. This study is performed by numerically
  integrating the planets and minor bodies in orbit around the parent
  star and in the Galactic potential. The calculations start 1\,Gyr
  ago and continue for 100\,Myr into the future. In this time frame,
  we simulate how asteroids (and planets) are ejected from the star's
  vicinity and settle in an Oort cloud and how they escape the local
  stellar gravity to form tidal steams.  A fraction of $0.0098$ to
  $0.026$ of the asteroids remain bound to their parent star.  The
  orbits of these asteroids isotropizes and circularizes due to the
  influence of the Galactic tidal field to eventually form an Oort
  cloud between $\sim 10^4$ and $\sim 2 \cdot 10^5$\,au.  We estimate
  that $\aplt 6$\% of the nearby stars may have a planet in its Oort
  cloud.  The majority of asteroids (and some of the planets) become
  unbound from the parent star to become free floating in the Galactic
  potential.  These \soli\, remain in a similar orbit around the
  Galactic center as their host star, forming dense streams of rogue
  interstellar asteroids and planets.

  The Solar system occasionally passes through such tidal streams,
  potentially giving rise to occasional close encounters with object
  in this stream. The two recently discovered objects, 1I/(2017 Q3)
  'Oumuamua and 2I/(2019 Q4) Borisov, may be such objects.  Although
  the direction from which an individual \solus\, originated cannot
  easily be traced back to the original host, multiple such objects
  coming from the same source might help to identify their origin.  At
  the moment the Solar system is in the bow or wake of the tidal
  stream of $\sim 10$ of the nearby stars which might contribute
  considerably to the interaction rate. Overall, we estimate that the
  local density of such left overs from the planet-formation process
  contribute to a local density of $1.2 \times 10^{14}$ per pc$^{-3}$,
  or $\apgt 0.1$ of the interstellar visitors originate from the
  obliterated debris disk of such nearby stars.  }

\maketitle

\section{Introduction}

The Sun is orbited by a cloud of minor bodies in a spherical
distribution between $\sim 10^4$\,au, and the Hill radius of the Solar
system in the Galactic potential. This cloud of objects, generally
referred to as the \cite{1950BAN....11...91O} cloud, may not be unique
to the Solar system: other stars may also have their own 'Oort' clouds
\citep{1989PhDT........14S}. The circumstances that lead to the Sun'
Oort cloud could equally well work for many nearby stars, in which
case, Oort clouds should be rather common.

Crucial to forming an Oort-like cloud is the presence of a debris disk
with relatively massive objects (which we call planets)
\citep{2013Icar..225...40B}. These planets perturb the low-mass minor
bodies (which we call asteroids) and drive their eccentricities to
high values ($\gg 0.9$) while preserving the closest approach distance
to the parent star. Upon each orbit, the asteroid is kicked into a
wider and more eccentric orbit while preserving pericenter
distance. Eventually, the Galaxy's tidal field, in which the planetary
system orbits, starts to circularize the orbit of the minor body
\citep{1987AJ.....94.1330D}. If the tidal field effectively reduces
the minor body's orbital eccentricity, it can remain bound to the
parent star and become a member of the local Oort cloud. If the body
becomes unbound, it continues as a free-floating asteroid, or \solus,
in the Galactic potential \citep[see][]{2019A&A...629A.139T}.

The recent arrival of two such \soli, 1I/(2017 Q3) 'Oumuamua
\citep{2017MPEC....U..181B,2017MPEC....U..183M,2017MPEC....W...52M}
and 2I/(2019 Q4) \cite{Borisov2019} \citep[and maybe even several
  others][]{2020P&SS..19004980F} in the Solar system makes it worth to
study the extent to which an evaporating Oort cloud or circumstellar
debris disk of nearby stars contribute to the frequency and
characteristics of such objects.

The two observed \soli\, were suggested to originate from the Oort
clouds of other stars
\citep{2018MNRAS.479L..17P,2018ApJ...856L...7R,2019AJ....157...86M},
or ejected from a proto-planetary disk \citep{2018ApJ...866..131M},
from a binary system \cite{2018MNRAS.478L..49J}, from a stellar
association \citep{2018ApJ...852L..27F}, or launched by as a tidal
fragment \citep{2020NatAs...4..852Z}.  At least 'Oumuamua and Borisov
do not seems to originate from the Sun's Oort cloud
\citep{2018DPS....5030101M,2020MNRAS.492..268H}.

Several studies have tried to identify the host
\citep{2018ApJ...852L..13Z,2020A&A...634A..14B}, but despite the
precision of the Gaia-DR2 astrometry \citep{2018A&A...616A...1G} and
the accurate ephemera for the two \soli, this remains
uneventful. Another complication in tracing back the interloper to its
host star is our lack of knowing the age of the object. Stars as well
as \soli\, move around in the Galaxy and tracing back one to another
can be arduous; much in the same way as it might be hard to identify a
dog owner in a dog-walking area.

Nevertheless, there is hope in identifying the \solus' host when
ejected asteroids linger around the source along it's Galactic
orbit. On the other hand, if \soli\, hang around the host star's
orbit, they may not appear at all to be launched from the star itself,
but from some distance away.  The asteroid may then appear to have
come from a somewhat different direction and with a different velocity
than the host star, making it hard to connect the two. The orbital
velocity of an asteroid in orbit around the parent star at the Hill
radius ($dv/v \equiv \delta v \aplt 0.1$\,km/s) is considerably
smaller than the orbit of the star around the Galactic center ($\sim
240$\,km/s). While the host star orbits in the Galactic potential, the
low-velocity ejected asteroids will form extended leading and trailing
arms along the hosts' orbit. We call this the proximity argument.

Rather than trying to identify the \solus' host star, we use the
proximity argument to determine the probable association between the
host star and its \soli. We do this by simulating the distribution of
asteroids that have become unbound from their parent star. Although
these asteroidal tails may be quite extended (several kpc), we limit
ourselves to the 200 stars nearest to the Sun (within about
16\,pc). We will assume that these stars had a disk of planets and
debris sometime in the past. In the simulations, Oort clouds form
through planets that eject asteroids. Most of these become unbound and
will orbit the Galaxy. We subsequently study the galactic distribution
of these ejected asteroids and investigate the probability of them
interacting closely with the Solar system.

We study the phase-space distribution of minor bodies from the moment
they enter the conveyor belt and are kicked out of the local
host-star's vicinity while sojourning the Milky way Galaxy. Our
calculations start by integrating the 200 nearby stars backward in
time in the Galactic potential for 1\,Gyr.  Once calculated backward,
they are provided with a planetary system and a population of minor
bodies as test particles in a disk. Each star with planets and minor
bodies is subsequently integrated forwards in time in the Galactic
potential until the current epoch, where they are observed today, near
the Sun. We call this the yo-yo scheme.  During this orbital
calculation we simulate the dynamical evolution of the planets and
minor bodies.  By that time, each star's Oort cloud has fully
developed, and the majority of minor bodies (and some of the planets)
are ejected on Galactic orbits. In this paper, we focus on the
distribution of these \soli\, in the vicinity of the Sun. For safety,
we continue the integration for another 100\,Myr after the present
time, in case the closest approach with the Sun happens to be in the
near future.

\section{Methods, initial conditions, and glossary}

This numerical study focuses on a hypothetical population of
interstellar asteroids that were liberated from their parent
stars. After liberation, they float freely in the Galactic
potential. Some of these \soli\, are welcomed as immigrants in the
Solar system or pass through the Oort cloud, but the majority will
just hover around orbiting the Galactic center without ever
interacting with the Solar system. The calculations depend on several
assumptions, and these are reflected, in part, by the initial
conditions chosen for this study. We realize that the reader may not
be satisfied with some of the choices made, and for this purpose, all
the run-scripts, the data, and picture-generating scripts to reproduce
this work are available through {\tt figshare}
\citep{SPZ2020_OCEII_Figshare} and github. The source code for the AMUSE
framework is available at {\tt https://github.com/amusecode/amuse}
\citep[see][]{portegies_zwart_simon_2018_1443252}.

\subsection{Glossary of terms}

The adopted terminology in this paper may be confusing at times.
Therefore a short glossary of terms used is presented.

\begin{itemize}
\item[] {\bf Asteroid:} low-mass object (compared to the planets) in a
  bound orbit around a star.
  
\item[] {\bf Conveyor belt:} Area in orbital parameter space (in
  particular semi-major axis and eccentricity) where an asteroid
  crosses the orbit of one or more of the major planets. These
  repeated perturbations induce small impulsive velocity-kicks causing
  the orbits to drift to higher eccentricity and larger semi-major
  axis while preserving pericenter distance (sometimes called
  eccentricity pumping).  The conveyor belt is indicated in
  figure\,\ref{fig:OCorbital_elements_init} as the area between the
  two black curves.

\item[] {\bf Frozen zone:} The area in parameter space (semi-major
  axis and eccentricity) where the orbits of asteroids are
  insignificantly perturbed by massive bodies (planets or stars) from
  shorter orbits, or stellar fly-by's.

\item[] {\bf Hill radius:} We define the Hill radius as the distance
  to the star for which the gravitational force is dominated by the
  star.  The Hill distance depends on the stellar mass and its orbit
  around the Galactic center. The Hill radius depends on direction,
  which results in an ellipsoid around the star. The force excerted by
  the star on objects within this volume exceeds the force from the
  rest of the Galaxy. The Roche radius
  \citep[][p.136]{1959cbs..book.....K} is defined as the radius of a
  sphere with the same volume as the Hill ellipsoid.

  In figure\,\ref{fig:OCorbital_elements_init} the Hill radius is
  presented as the solid red curve, here at roughly $2 \cdot
  10^5$\,au.  In our calculations, the Hill ellipsoid only considered
  in post processing the data. During the calculations weather or not
  an asteroid (or planet) is bound to a star is calculated by
  integrating the equatios of motion of the star, planets, and
  asteroids in the potential of the Galaxy. In
  figure\,\ref{fig:RHill_in_Galaxy_with_OC} we present the
  (post-processing) Hill radius for all stars in our simulation to
  illustrate its variation over time.

\item[] {\bf Kuiper cliff:} The rather sudden drop in the density of
asteroids at the far side of the Edgewordt-Kuiper belt
\citep{1943JBAA...53..181E,1951astr.conf..357K}.

\item[] {\bf Oort cloud:} The Oort cloud is reserved for bound objects
  but for which a 1\,\MSun\, body passing on a hyperbolic orbit at the
  Hill radius distance induces a relative velocity-perturbation of
  $\delta v = 10^{-5}$.  Here we define the relative velocity
  perturbation $\delta v \equiv |v_\odot-v_{\rm Gal}|/v_\odot$ as the
  change in velocity of an object at apocenter in its orbit around a
  star due to the presence galactic tidal field
  \citep[see][]{SPZJilkova2015}.  The inner edge of the Oort cloud is
  presented in figure\,\ref{fig:OCorbital_elements_init} by the
  right-most cyan curve.
  
\item[] {\bf Parking zone:} The region between the Kuiper cliff and
  the Oort cloud. The parking zone in the Solar system is rather
  empty.

\item[] {\bf Rogue planet:} Planet that is unbound from any star.
  Note that strictly speaking, the term free-floating is misleading
  because they are still bound to the Galaxy.
  
\item[] {\bf \Solus\, (plural: \soli):} Asteroid or comet that is not
  bound to any star but floating freely in the Galaxy.  Asteroids to
  the far side of the Hill radius are generally not bound to the
  parent star, but have become a member of the Galaxy.

\end{itemize}

\subsection{The simulation environment}

The calculations in this study are performed using the Astrophysics
Multipurpose Software Environment \citep{2018araa.book.....P}. AMUSE
is a modular language-independent framework for homogeneously
interconnecting a wide variety of astrophysical simulation codes.  It
is built on public community codes that solves gravitational dynamics,
hydrodynamics, stellar evolution, and radiative transport.  The
underlying codes are written in high-performance compiled languages
that do not require recompilation when combined with Python
scripts. The framework adopts {\em Noah's Arc} philosophy, in which
there are at least two codes that solve for the same physics
\citep{2009NewA...14..369P}.

Most calculations are carried out using a combination of symplectic
direct N-body, test-particle integration, and the tidal field of the
Galaxy. For the former two, we use {\tt Huayno}, which adopts a
recursive Hamiltonian-splitting strategy \citep[much like {\tt
    bridge}, see][]{2007PASJ...59.1095F} to generate a symplectic
integrator that conserves momentum to machine precision
\citep{2012NewA...17..711P}. For the asteroids, we employed the hold
drift-kick-drift scheme in the integrator with a time-step parameter
of 0.02 (this results in an integration time-step of about
10-days). We use the GPU enabled version, which employs the {\tt
  Kirin} GPU library
\citep{2007NewA...12..641P,2014ascl.soft01001B}. The direct N-body
code and the Galactic potential are coupled using the hierarchical
non-intrusive code-coupling strategy discussed in
\cite{ZWART2020105240} with an interaction time-step of 100\,yr.

We ignore stellar mass loss in this study. The most massive star in
our sample is 1.23\,\MSun\, (HD 346704), which will not leave the
main-sequence within $\sim 5.9$\,Gyr (for Solar metallicity). This is
considerably longer than the time frame over which this study is
conducted.

\subsection{Reconstructing the debris distribution near the Sun}

\subsubsection{The yo-yo approach}

This study strives to achieve a representative distribution of
interstellar asteroids in the Solar system's vicinity. The yo-yo
scheme is carried out in two steps: In the first step (yo), we
integrate the known nearby stars backward through time in a
representative Galactic potential. In the second step (yo), we
integrate these same stars from their back-calculated location in the
Galaxy forwards in time to the current epoch. In the second step, each
star will have a disk of planets and asteroids which are also
integrated in the Galactic potential.

In the next sections, we discuss the following steps in the yo-yo
procedure.
\begin{itemize}
\item[A:] Select the 200 nearby stars used in this study (see Sect.\,\ref{Sect:IC_A}).
\item[B:] Calculate their Galactic location 1\,Gyr
ago (see Sect.\,\ref{Sect:IC_B}).
\item[C:] Supply each star with planets and asteroids (see Sect.\,\ref{Sect:IC_C}).
\item[D:] Calculate star, planets, and asteroids forwards in time for
1.1\,Gyr in the Galactic potential (see Sect.\,\ref{Sect:IC_D}).
\end{itemize}

\subsubsection{Selecting the 200 nearest stars}\label{Sect:IC_A}

We start by selecting the 200 stars currently nearest to the Sun. The
proper motions, radial velocities, positions, and masses of these
stars are taken from the catalog of
\cite{2019A&A...629A.139T,2019yCat..36290139T}. The proper motion,
distance, and position on the sky in this catalog are taken directly
from Gaia DR2 \citep{2018A&A...616A...1G,2018A&A...616A...8A}. The
radial velocities are derived by cross-matching Gaia DR2 with RAVE-DR5
\citep{2017AJ....153...75K}, GALAH DR2 \citep{2018MNRAS.478.4513B},
LAMOST DR3 \citep{2012RAA....12..723Z}, APOGEE DR14
\citep{2018ApJS..235...42A}, and XHIP
\citep{2012AstL...38..331A}. Only stars with a relative uncertainty of
$< 20$\% on the parallax are selected. Regretfully, this excludes some
interesting stars, such as Glise\,876, which, according to
\cite{2018A&A...610L..11D} is a favorable candidate for the origin of
'Oumuamua.

After selecting the stars, their coordinates on the sky, distances,
proper motions and radial velocities are converted to galactic
position and velocity in Cartesian coordinates, assuming the solar
position of $(-8.34, 0.0, 0.027)$\,kpc with a velocity of $(11.1, 240,
7.25)$\,km/s.

In figure\,\ref{fig:local_potential_xy} we present a projection in the
Galactic plane of the relative equipotential-surface of these
stars. The red circle at $\sim 16$\,pc indicates the distance to the
nearest 200 stars, which form the basis of our calculations.

\subsubsection{Reconstructing the stellar birth locations}\label{Sect:IC_B}

As the first step in generating initial conditions, we integrate the 200
selected stars backward in time in the Galactic potential. We use
the model of the Galaxy by
\cite{2015MNRAS.446..823M}. 

This back-integration is performed for 1\,Gyr. Some stars may be
considerably older than that, and a few may even be younger. Therefore, these
initial locations should be considered approximate and are
meant to reconstruct the past several hundred Myr of the 
stars' evolution, rather than the full evolution since the Sun's birth. More extended
integration of the Galactic stars would eventually require us
to take the Galaxy's dynamical evolution into account. In our
calculations, however, we adopted a semi-analytic dynamical model of
the Galactic potential, which represents the current epoch, but which
can probably not be extrapolated backward reliably beyond a Gyr.

After calculating each star backward in time (for a Gyr) in the
Galactic potential they arrive at what we consider their initial
positions and velocities. In
figure\,\ref{fig:solar_orbit_in_the_galaxy_at_0Gyr} we present
projected views of the Galaxy in the three fundamental Cartesian
planes and a small close-up at the Sun's location in the Galactic disk
in the X-Y plane (top right). At the start of our simulations (1\,Gyr
ago) none of the 200 stars current in the vicinity of the Sun were
within 6\,pc (see
figure\,\ref{fig:solar_orbit_in_the_galaxy_at_0Gyr}).

The back-calculations result in a unique position and velocity for
each star, and by calculating them forwards again, they arrive at
precisely their starting points; the orbital integration of the stars
in the semi-analytic Galactic potential is not chaotic: but see the
discussion in section\,\ref{Sect:discussion}.

\begin{figure}
\centering
\includegraphics[width=0.8\columnwidth]{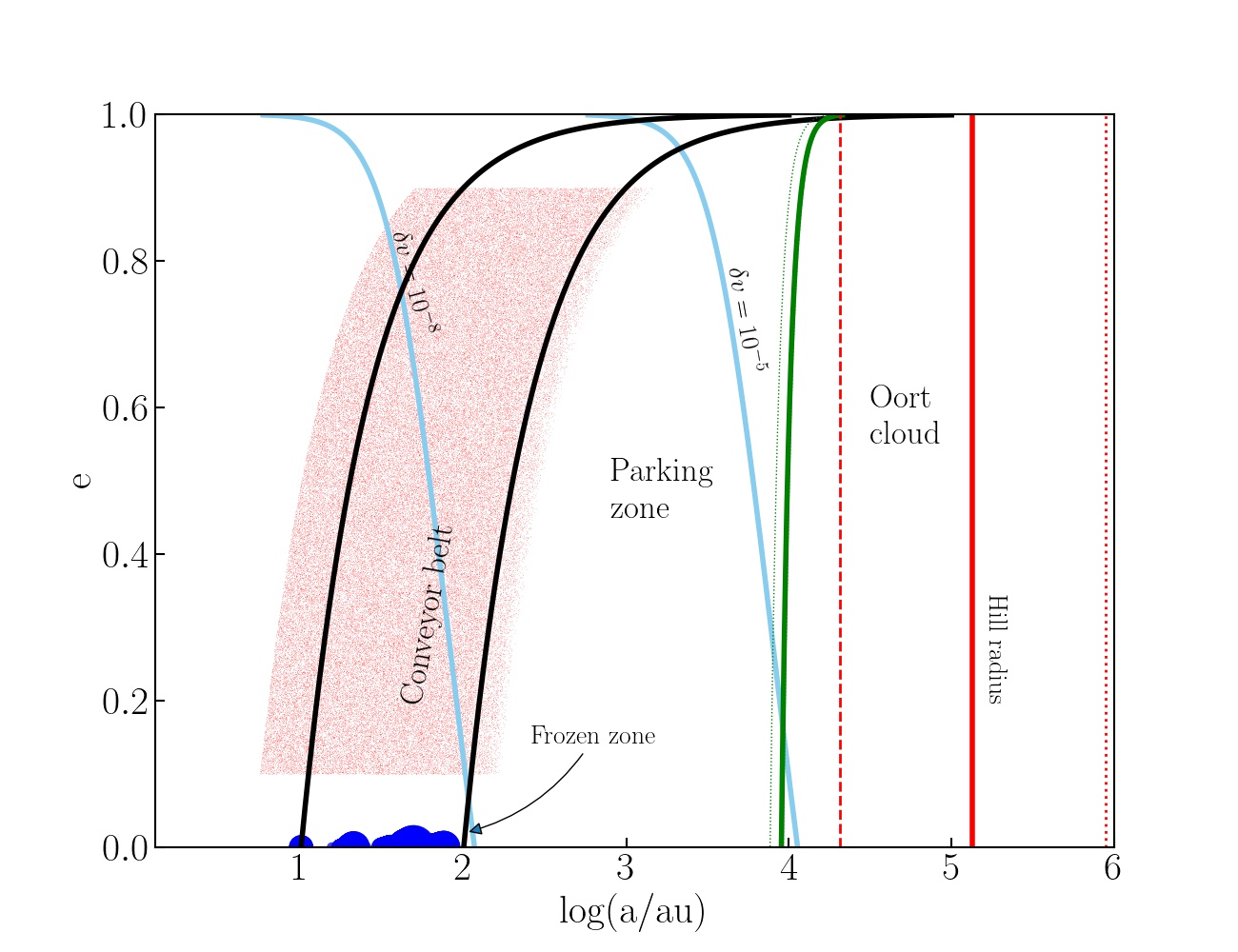}
\caption{Orbital parameters of planets (blue bullets) and 141864
  asteroids (red points) around the 200 nearest stars 1\,Gyr ago. Here
  the right-most light blue curve gives the approximate inner edge of
  the Oort cloud using a mean local stellar mass of $\sim
  1.0$\,\MSun\, and a mean distance according to the Sun's Hill radius
  in the Galatic potential ($\sim 0.65$\,pc, red vertical line).  The
  vertical dashed and dotted curves give the smallest and largest Hill
  radii experienced by any star throughout the simulations.  The inner
  light-blue curve is given for a relative perturbation of $\delta v =
  10^{-8}$, the outer light-blue curve gives $\delta v = 10^{-5}$. The
  green curve gives the distance from the host star for which the
  circularization time-scale of the Galactic tidal field is comparable
  to the eccentricity pumping time-scale of a test particle in the
  conveyor belt. The orbital configuration at today $1$\,Gyr after the
  start of the simulation, is presented in
  figure\,\ref{fig:OCorbital_elements_init}.
This figure is generated with the Python script {\tt
plot\_semimajor\_axis\_vs\_eccentricity\_of\_planets\_and\_comets.py}. This
script, the required data, and the scripts to generate any of the
figures in this manuscript including all required data are available
through {\tt figshare} \citep{SPZ2020_OCEII_Figshare}.  An animation
of the simulation is available at {\tt https://youtu.be/0fYeAW3e9bQ}.
}
\label{fig:OCorbital_elements_init}
\end{figure}

\begin{figure}
\centering
\includegraphics[width=\columnwidth]{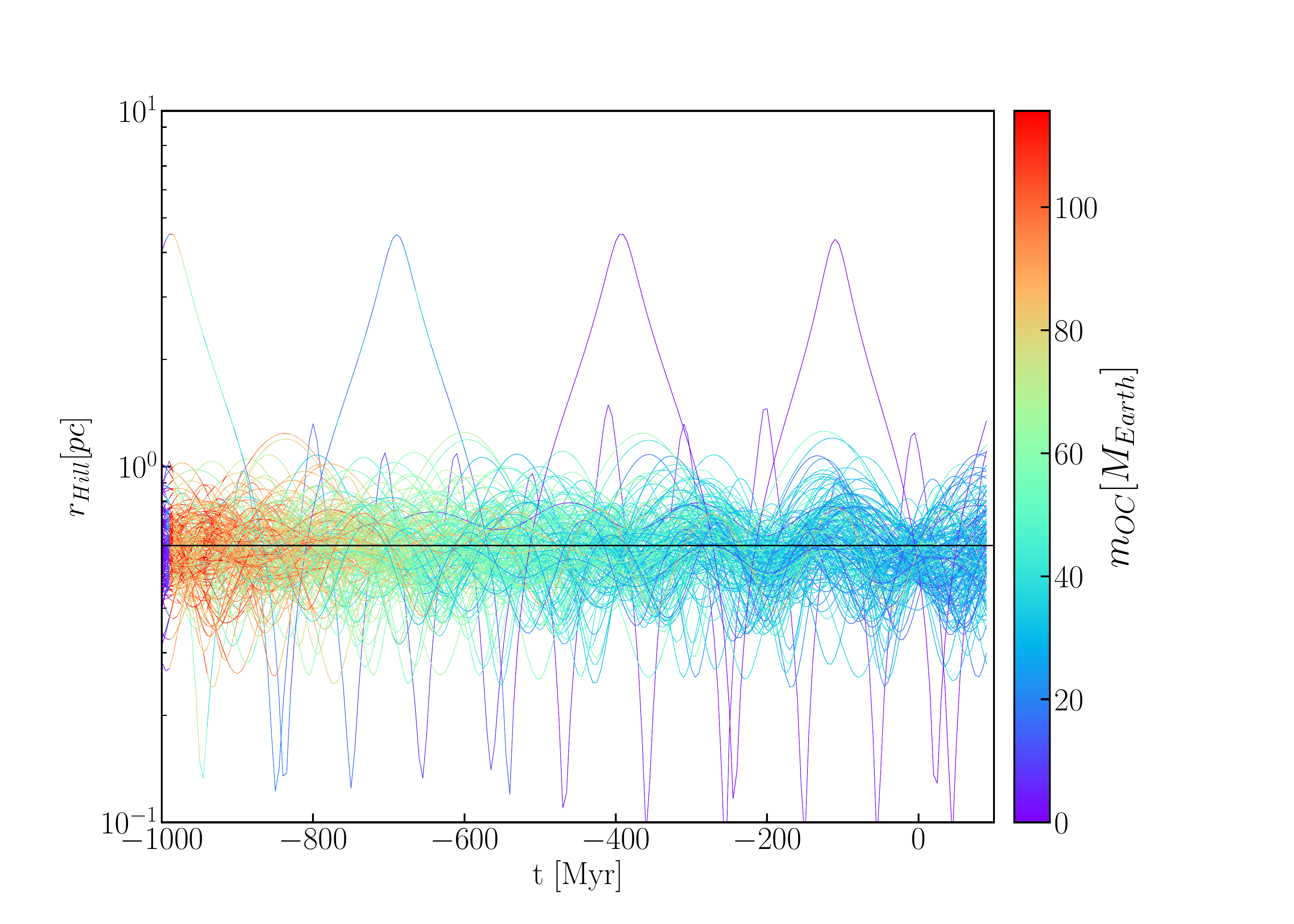}\\
\caption{Approximate size of the Hill radius as a function of time for
  all the stars in the simulation since their birth until today and
  100\,Myr into the future. The mass of the Oort cloud is color coded
  (legend to the right).  The star HD\,103095 (see also
  figure\,\ref{fig:galactic_orbit_of_specific_stars_HD}) jumps out
  again with its curious orbit.  }
\label{fig:RHill_in_Galaxy_with_OC}
\end{figure}

\begin{figure}
\centering
\includegraphics[width=\columnwidth]{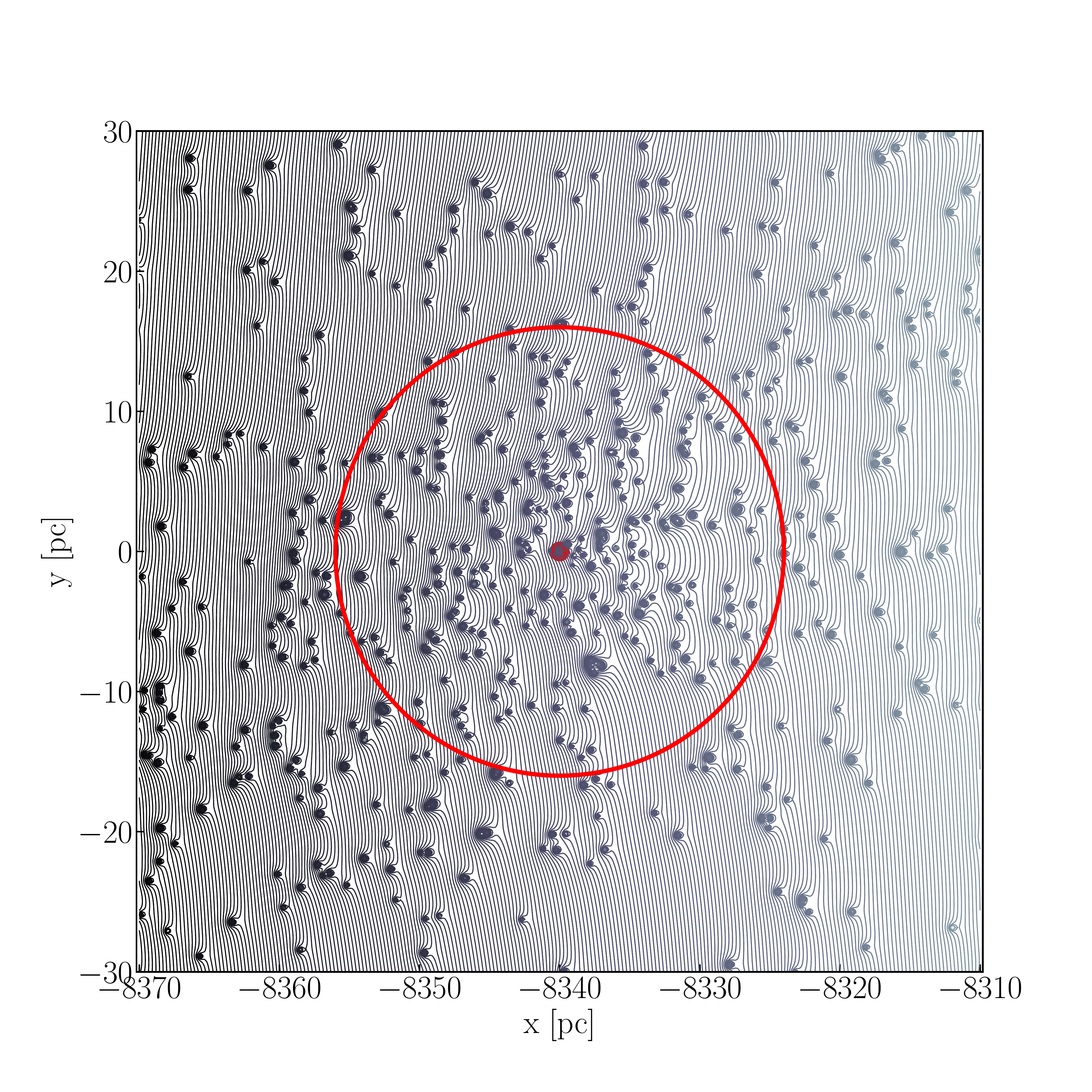} \\
\caption{Equipotential surface of $\sim 795$ stars within 50\,pc of
the Sun projected on the xy-plane in the Galactic potential. To make
the potential more visible, we magnified the stellar masses by a
factor 300. The Sun is indicated with the red dot in the middle of
the figure. The red circle, at $\sim 16$\,pc indicates the inner
region for which the analysis in this paper is performed. This
volume contains the 200 nearest stars \citep{2019yCat..36290139T}.
The script to generate this image, {\tt
plot\_nearby\_stars\_equipotential\_surface.py}, and the required
data (filename {\tt gaia\_nearby\_stars.amuse}), are both available
at figshare \citep{SPZ2020_OCEII_Figshare}. }
\label{fig:local_potential_xy}
\end{figure}

\begin{figure*}
\centering
\includegraphics[draft=false,width=\linewidth]{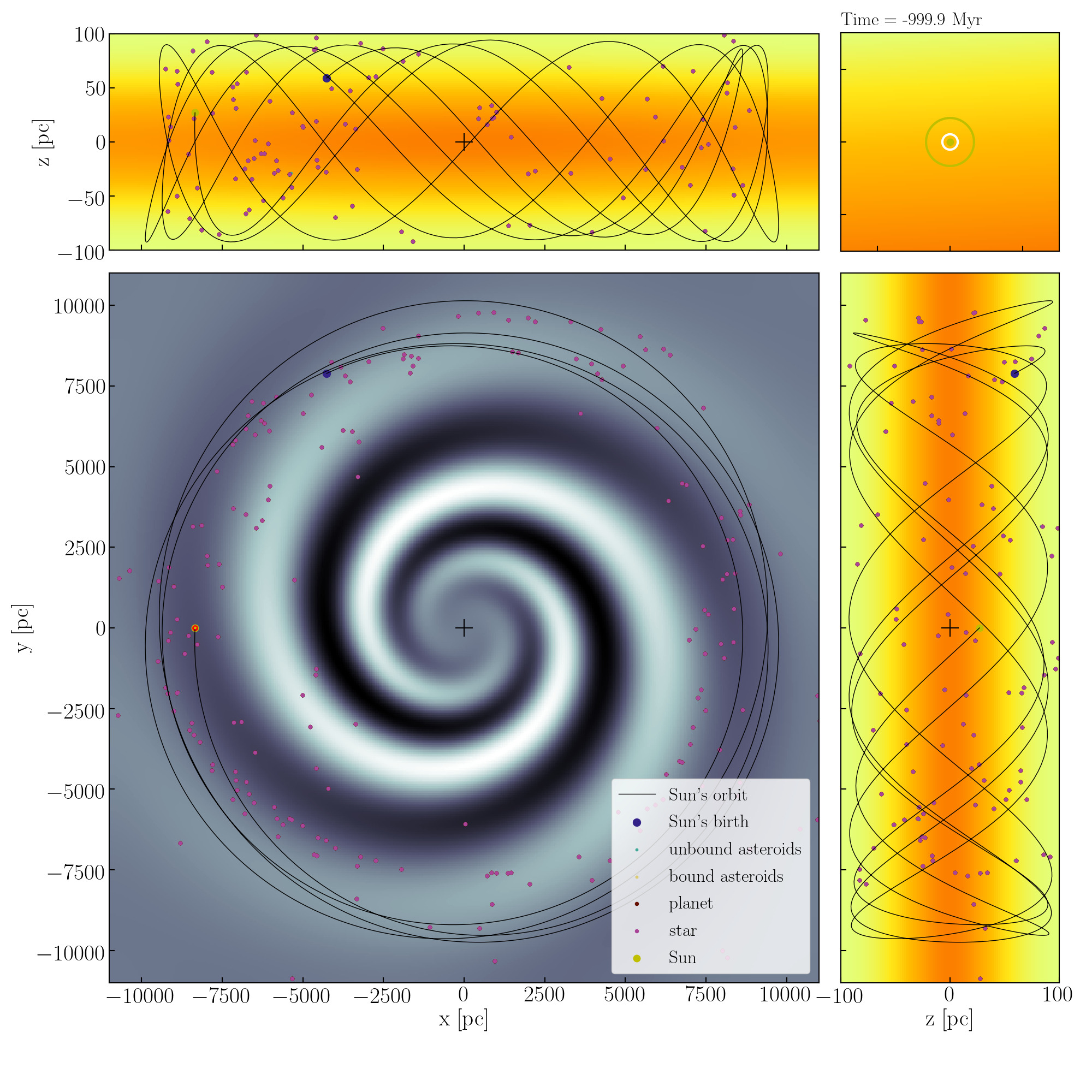}\\
\caption{Positions of the 200 stars at their birth location in the
  potential of the Galactic 1.0\,Gyr before the present. The solid
  curve shows the orbit of the Sun for the last Gyr. The colored
  background represents the adopted Galaxy potential: bottom left, top
  left, and bottom right views show the various Cartesian
  coordinates. The spiral structure is visible in the $X$-$Y$- plane
  (lower left), whereas the other two give panels only the edge-on
  views of the Galaxy is visible. The Sun's starting position is
  indicated with the black bullet, it's end position with a yellow
  bullet.  The planets and asteroids of all the nearby stars overlay
  at this initial snapshot because the planetary systems are smaller
  than the symbol-size in the panels.  The top-right corner-panel
  shows a magnified view of 6 by 6 parsec around the Sun. The outer
  circle shows the Hill radius at $\sim 0.65$\,pc, the inner circle
  represents the inner edge of the Oort cloud at about $\sim
  0.05$\,pc.  In this initial image, none of the stars are within
  6\,pc of the Solar system.  The color graident in the top-right
  panel shows the local variation in the Galactic potential.  
}
\label{fig:solar_orbit_in_the_galaxy_at_0Gyr}
\end{figure*}

\subsubsection{Initialization of planetary systems with asteroids}
\label{Sect:IC_C}

Once we have determined each star's position and velocity a Gyr ago
(see figure\,\ref{fig:solar_orbit_in_the_galaxy_at_0Gyr}), planets,
and asteroids are added. The planets are generated using the
oligarchic growth model
\citep{1998Icar..131..171K,0004-637X-807-2-157} using the
implementation of \cite{0004-637X-775-1-53} We assumed the planets to
form from a disk with 1 percent of the stellar mass with a minimum
distance of 10\,au from the host star, and adopting an outer disk
radius of 100\,au. The oligarchic growth model then leads to 3 or 4
giant planets, dependent on the disk's mass.

\begin{figure}
\centering
\includegraphics[width=\columnwidth]{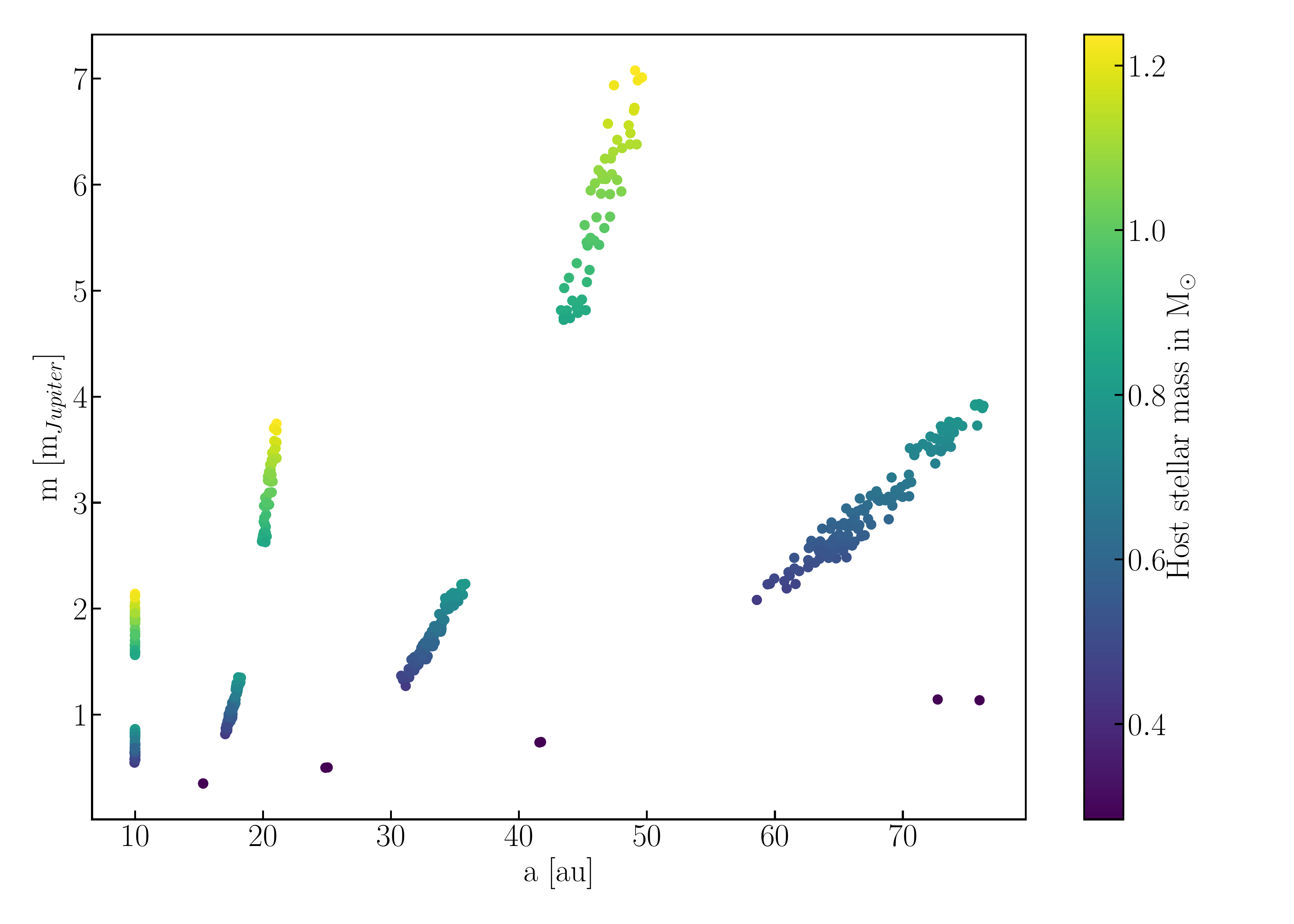}\\
\caption{Initial planet mass as a function of the orbital separation
  around their parent star. The mass of the host star is color coded
  (see the color bar to the right). The two regimes, relatively
  low-mass ($\aplt 0.85$\,\MSun) acquiring 4 planets according to the
  olycharcig growth model, whereas more massive stars only receive
  three planets.  }
\label{fig:initial_semimajor_axis_and_planet_mass}
\end{figure}

In figure\,\ref{fig:initial_semimajor_axis_and_planet_mass} we present
for each star, the masses of the planets, and their distance to the
host star. The inner-most planet is always initialized at 10\,au, the
other planets follow the oligarchic growth model, resulting in
circular orbits in the same plane. The outer-most planet then arrives
at $60$ to $76.3$\,au (see
figure\,\ref{fig:initial_semimajor_axis_and_planet_mass}). Planet
masses range from $0.35$\,M$_{\rm Jupiter}$ to $7.08$\,M$_{\rm
  Jupiter}$.

\begin{figure}
\centering
\includegraphics[width=\columnwidth]{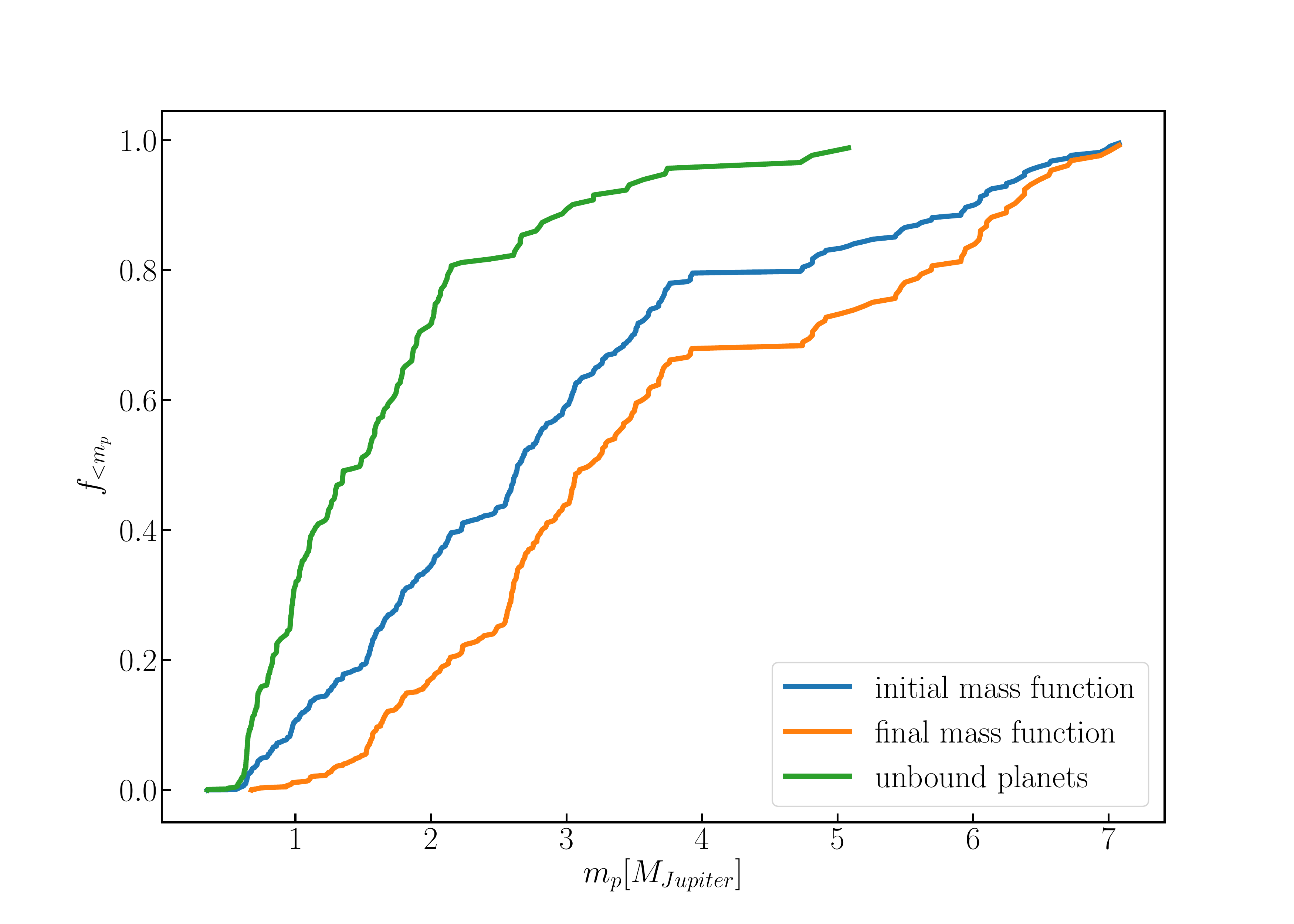}\\
\caption{Cumulative mass-distribution of the planets at the start of
the simulation (blue) and at the current epoch ($1$\,Gyr after the
start of the simulations), in orange (for the bound planets) and
green (for the unbound planets).}
\label{fig:initial_planet_mass}
\end{figure}

Several regimes are visible in
figure\,\ref{fig:initial_semimajor_axis_and_planet_mass}. These result from the Oligarchic model in which planet masses increase
with distance to the host star. Stars more massive than about
0.85\,\MSun\, tend to acquire three relatively massive planets, whereas
the lower mass stars receive four planets of somewhat lower mass. The
cumulative distribution of the initial planet masses is presented in
figure\,\ref{fig:initial_planet_mass} (blue curve).

Once the planets are initialized, we add a population of asteroids as
test particles in the planets' plane. Each star receives 1000 minor
bodies per 1\,\MSun\, of the stellar mass. The pericenter distance was
selected homogeneously between 5\,au and 150\,au with a random
eccentricity between 0.1 and 0.9.  This choice prevents asteroids to
settle in resonances with the giant planets, where they continue to
consume computer resources which is not the focus of this study.

The conditions for asteroids are presented in
figure\,\ref{fig:OCorbital_elements_init}, where each red dot
represents an asteroid. This choice of relatively high eccentricities
may sound a bit odd for a debris disk, but it effectively places the
asteroids in the planetary conveyor belt. In this parameter-space
region, asteroids receive repeated gravitational kicks from the
planets, driving them into wider and more eccentric orbits while
preserving pericenter distance.

Before we start integrating each planetary system in the Galactic
potential, the orbital plane is rotated to a random orientation
(isotropically oriented in all fundamental angles). Each star with
planets and asteroids then starts its orbit 1\,Gyr ago at the
calculated position in the Galaxy.

Each star with planets and asteroids are considered isolated with respect to the other stars. Since they are born throughout a wide area
in the Galaxy (see
figure\,\ref{fig:solar_orbit_in_the_galaxy_at_0Gyr}) this assumption
seems valid. Even today (when reaching the highest congregation) the
stars are sufficiently far apart that the mutual influence of one star
on the planetary system of another star is negligible. This is also
illustrated in the top-right panel in
figure\,\ref{fig:solar_orbit_in_the_galaxy_at_0Gyr}, which shows
today's vicinity of the Sun.

While generating the planetary systems' initial conditions, we ignore
the binarity (or higher-order multiplicity) of some of the nearby
stars. Several nearby stars are known to be multiple, and one of the
important contributors to the local population of \soli, 61~Cyg~A and
B, actually is a binary system. We also ignored the known planets
orbiting some of the catalog  stars. This poses less of a limitation to
the starting conditions because the known planets tend to have much
tighter orbits than we adopted here, and they hardly provide
constraints on the wider orbits. We leave the improvement of these
assumptions for future work.

In figure\,\ref{fig:RHill_in_Galaxy_with_OC} we show, an a function of
time, as estimate for the Hill radius for each star in the simulation.
At the earliest phase the Oort clouds are empty, which shows-up as the
dark blue colors for all stars. But after some 100\,Myr the respective
Oort clouds start to be populated. As time progresses, these Oort
clouds are depleted.  The size of the Hill radius varies with the
distance to the Galactic center. Near the Galactic center the Hill
radius is small, whereas it grows when the star approaches apocenter
in its orbit around the Galactic center.  A smaller Hill radius,
approaching perigalacticon, leads to enhanced mass loss from the Oort
cloud.

\begin{figure}
\centering
\includegraphics[width=\columnwidth]{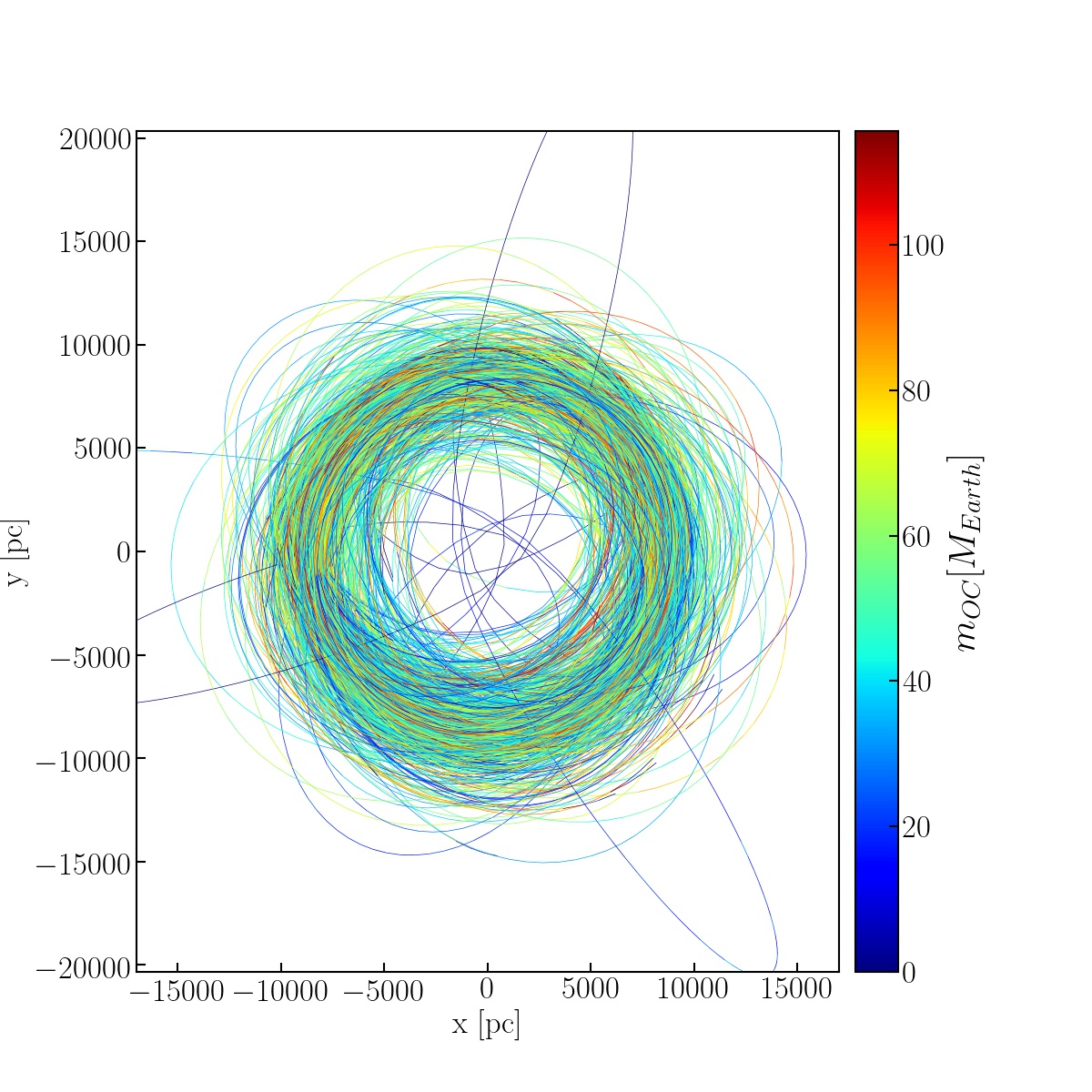}\\
\caption{Mass of the Oort cloud for all the stars in the simulation
  since their birth until today and 100\,Myr into the future. The
  legend, to the right, give the color coding in units of Earth
  masses. The star HD\,103095 (see also
  figure\,\ref{fig:galactic_orbit_of_specific_stars_HD}) jumps out as
  having the most elliptical orbit.}
\label{fig:Orbit_in_Galaxy_with_OC}
\end{figure}

To illustrate the orbits of the stars, we also present a projected
view of the Galaxy with the orbits all stars in
figure\,\ref{fig:Orbit_in_Galaxy_with_OC}. Here we adopt the same
color coding for the stellar orbits to make the comparison with
figure\,\ref{fig:RHill_in_Galaxy_with_OC}.

\begin{figure*}
\centering
\begin{tabular}{c}
\includegraphics[width=\columnwidth]{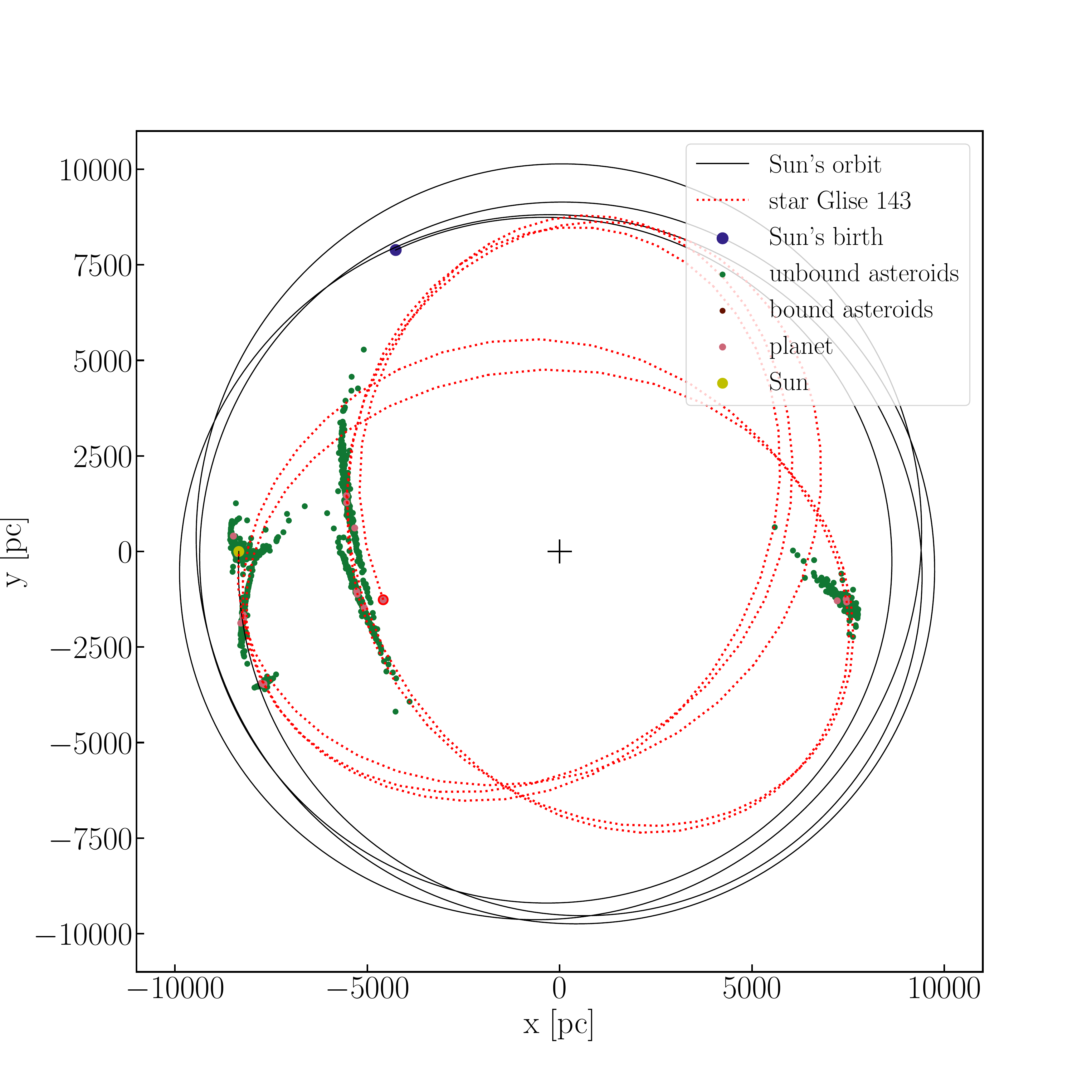}
~\includegraphics[width=\columnwidth]{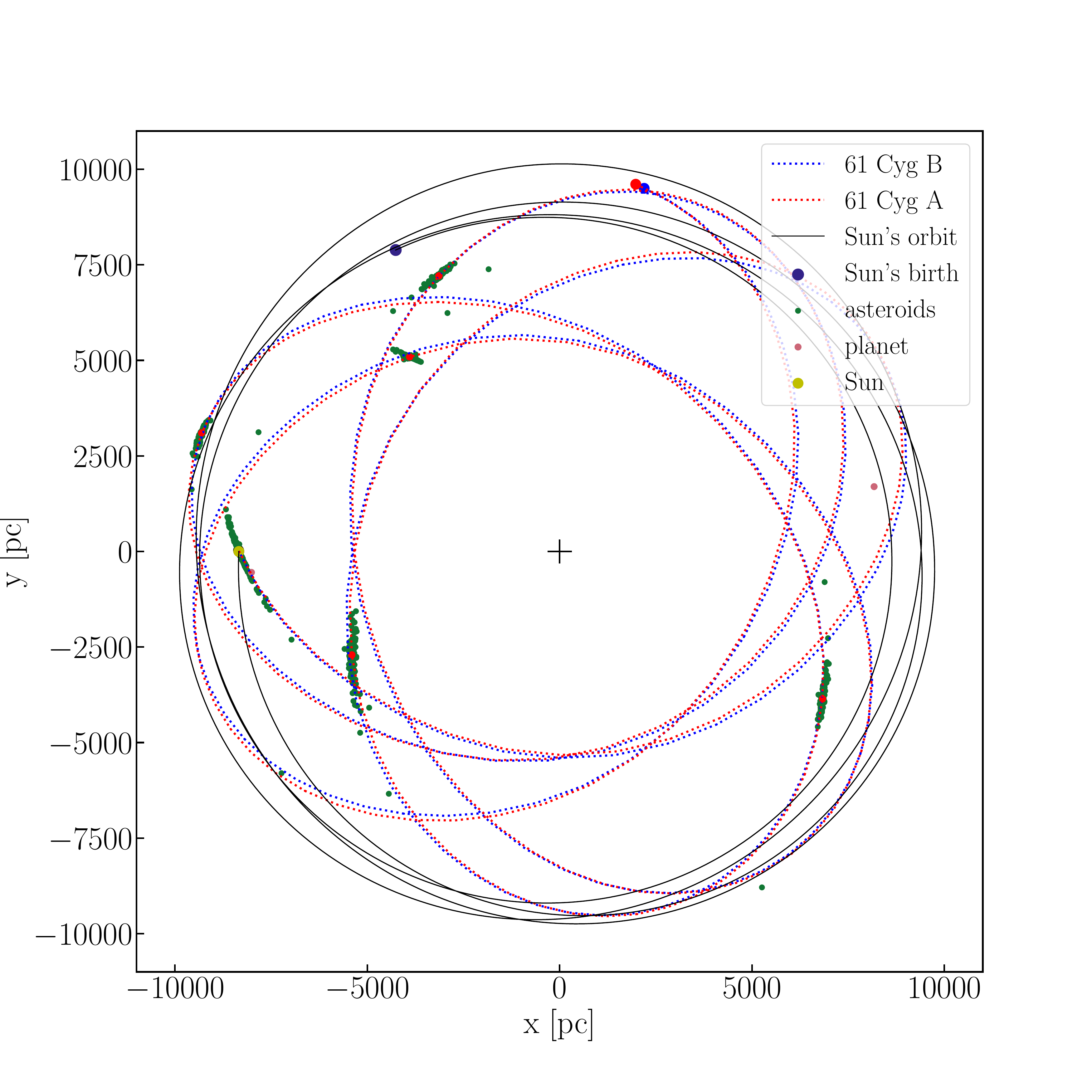}
\end{tabular} 
\caption{Examples of the Galactic orbits of Glise\, 143 (left panel)
  and the 61\, Cyg binary (right). For Glise\, 143 and 61\, Cyg\, A we
  plot the position of the star and asteroids 1.0\,Gyr ago, 0.8, 0.6,
  0.4, 0.2\,Gyr ago and today. We plotted the orbits of both stars
  61\, Cyg\, A (red dotted line) and 61\, Cyg\, B (blue dotted
  line). Both stars start (1\,Gyr) ago at the big bullet point to the
  top of the figure. The planets for 61\,Cyg\,A are plotted only for
  the current epoch (when the star is near the Sun, to the left). At
  this time, two planets still orbit the star, but the other two
  planets were ejected early in the calculation. By the current time
  the ejected planets have moved away consideraby from the star. The
  main differences in the Galactic orbits of the two stars 61\, Cyg\,
  A and 61\, Cyg\, B (compare the red and blue dotted curves in the
  right-hand panel) is caused by the ejected of these two planets. }
\label{fig:galactic_orbit_of_specific_stars}
\end{figure*}

\begin{figure*}
\centering
\includegraphics[width=\linewidth]{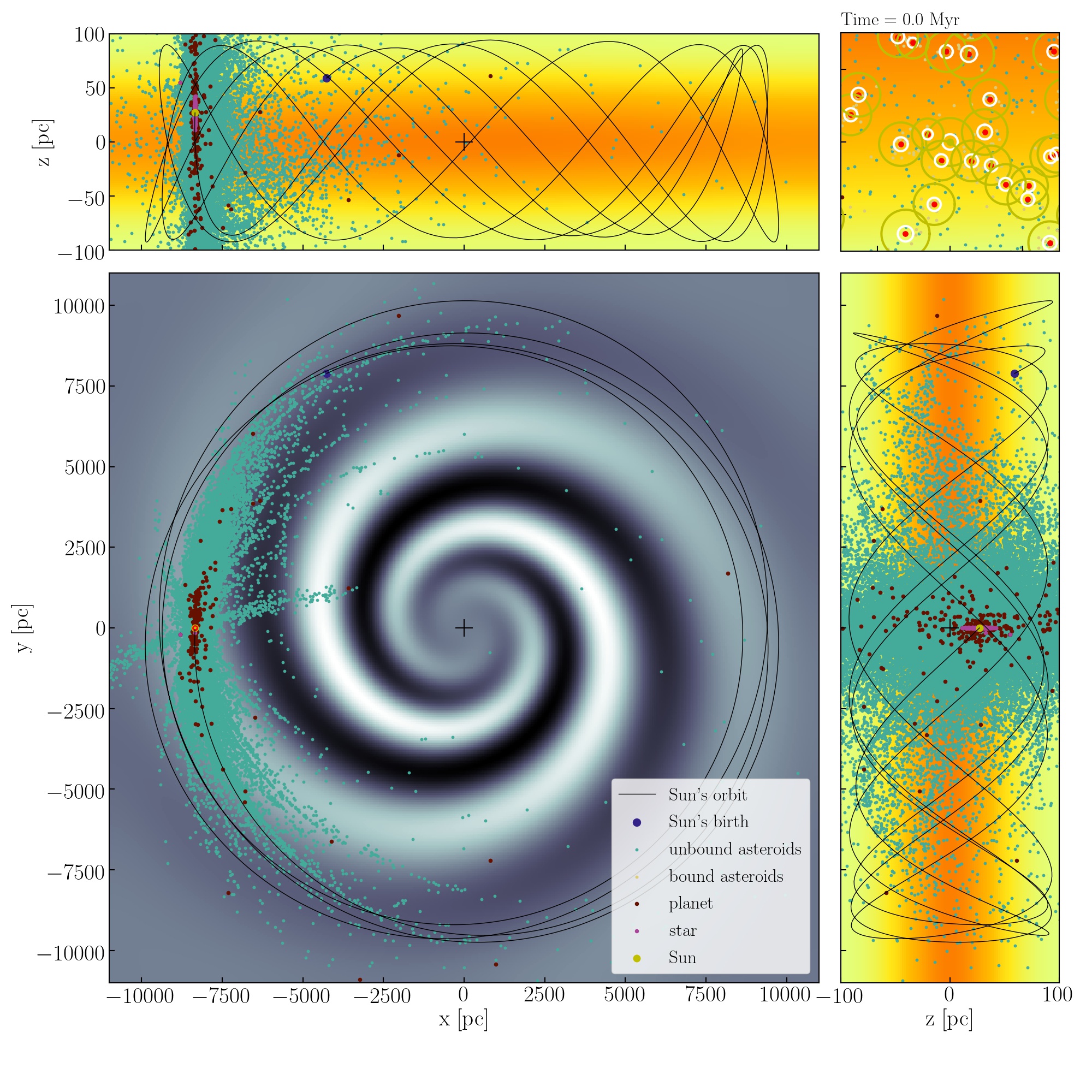} \\
\caption{Current view of the Mikly way Galaxy along the three major
  projections and a close-up centered around the Sun in the top-right
  corner (as in
  figure\,\ref{fig:solar_orbit_in_the_galaxy_at_0Gyr}. The orange,
  green, and red points indicate the bound and unbound asteroids and
  the planets, respectively. The stars are presented in bright red
  points.  The Sun's orbit in the Galactic potential is presented as
  the thin black curve which starts at the black bullet and ends at
  the yellow dot (to the left).  For an animation, see
  {\tt https://youtu.be/0fYeAW3e9bQ}.  }
\label{fig:solar_orbit_in_the_galaxy_at_1Gyr}
\end{figure*}

\begin{figure}
\centering
\includegraphics[width=\columnwidth]{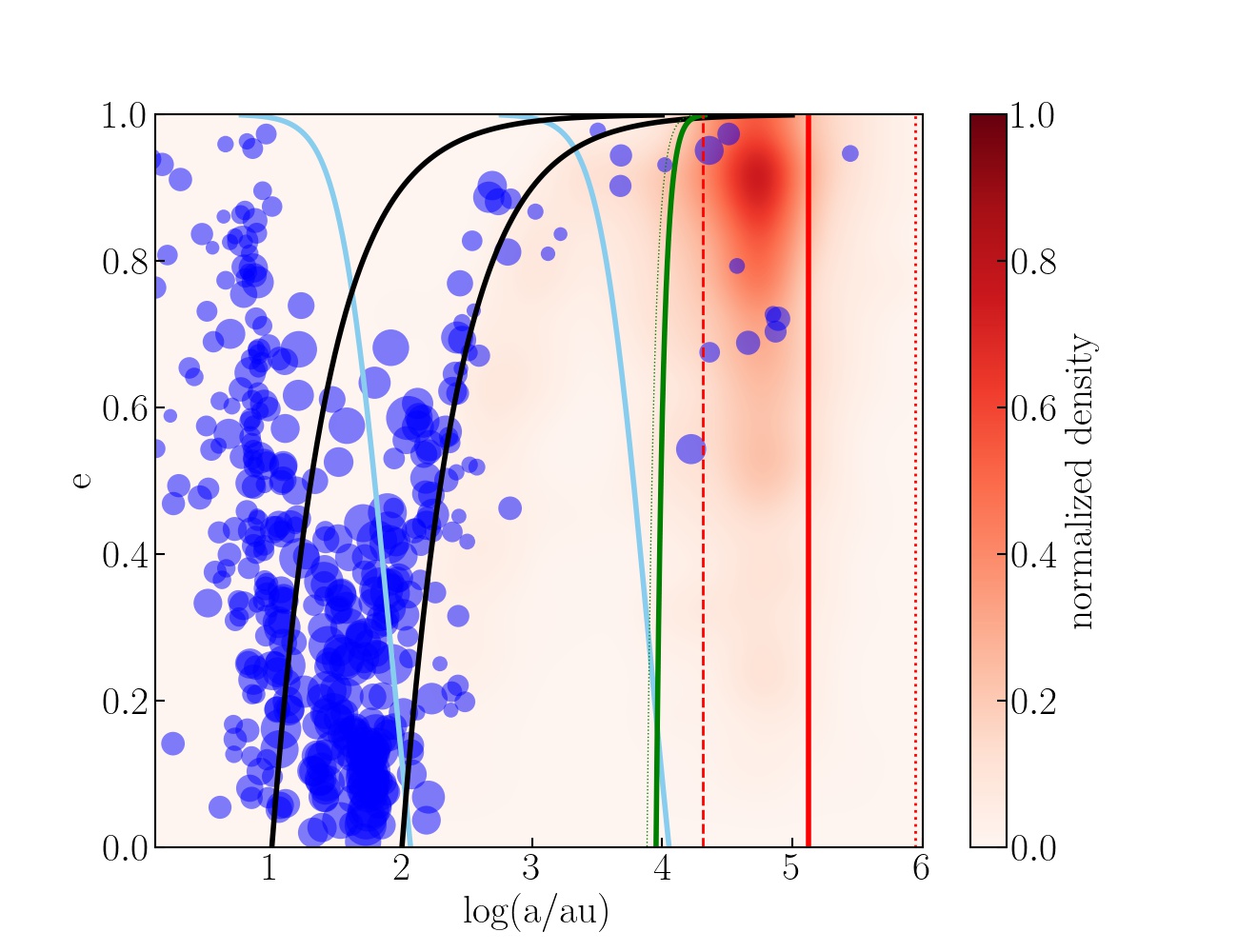}
\caption{Orbital parameters of planets (blue bullets) and 141864
  asteroids (red fuzz) at 1\,Gyr after the start of the simulation.
  The various curves are described in
  figure\,\ref{fig:OCorbital_elements_init}.  The size of the blue
  bullets is proportional to the mass of the planet. The two black
  curves indicate the pericenter distances associated with these
  mimimun and maximum initial orbital separations, indicating the
  limits of the conveyor belt. The minor bodies are presented as small
  red bullet points. In figure\,\ref{fig:OCorbital_elements_today} we
  presented the initial conditions, and explained the terminology. }
\label{fig:OCorbital_elements_today}
\end{figure}

\begin{figure}
\centering
\includegraphics[width=\columnwidth]{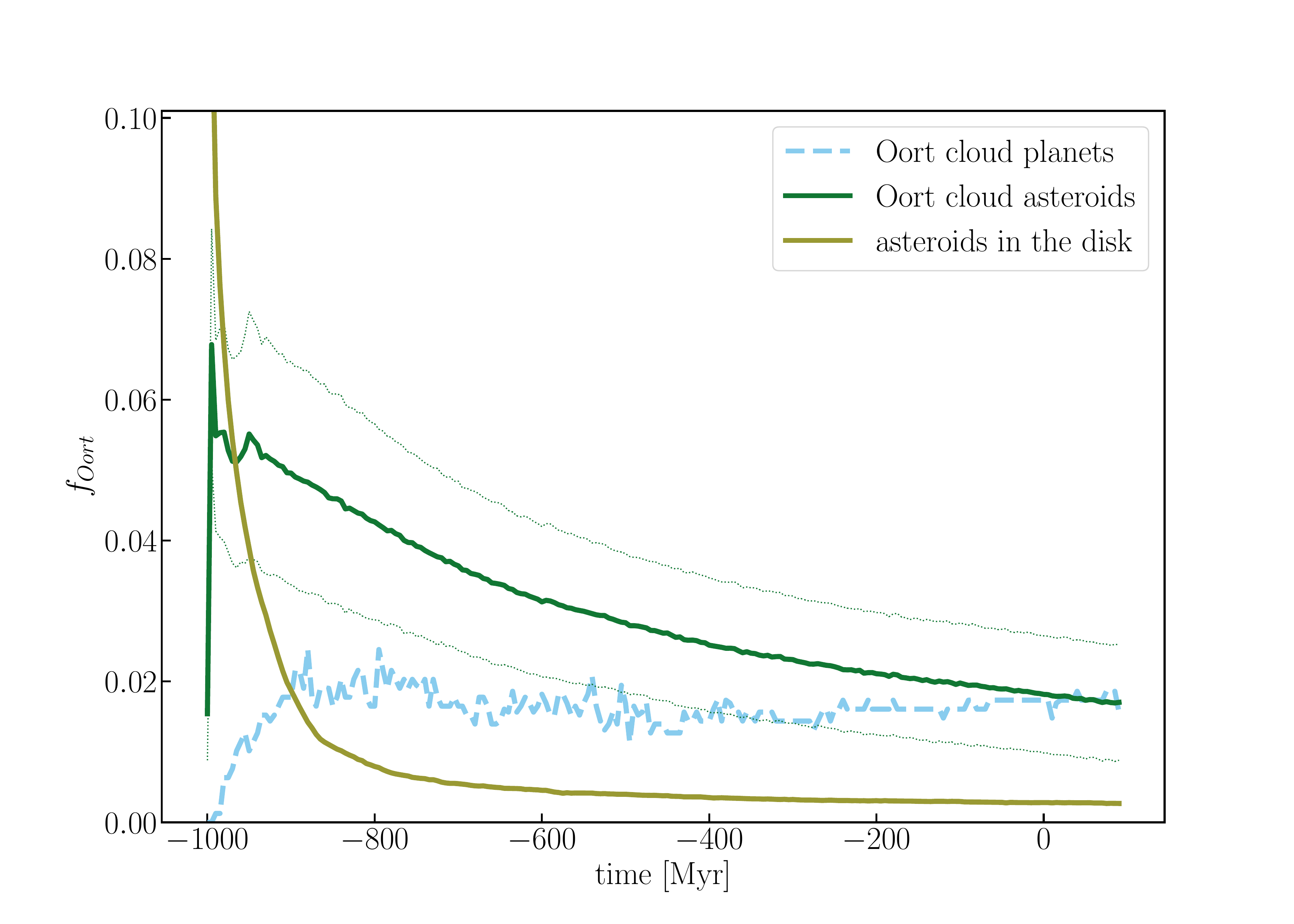}\\
\caption{Evolution of the fraction of mass in a bound Oort cloud of
their respective stars (solid green curve). The dotted curves give
the dispersion (one standard deviation) among stars. The blue dashed
curve gives the fraction of planets that populate the Oort
cloud. The ocher-colored curve gives the fraction that remains in
the disk either in resonant orbits or in the parking zone (outside
the perturbing influence of the planets).}
\label{fig:Oortcloud_fraction}
\end{figure}

\begin{figure}
\centering
\includegraphics[width=\columnwidth]{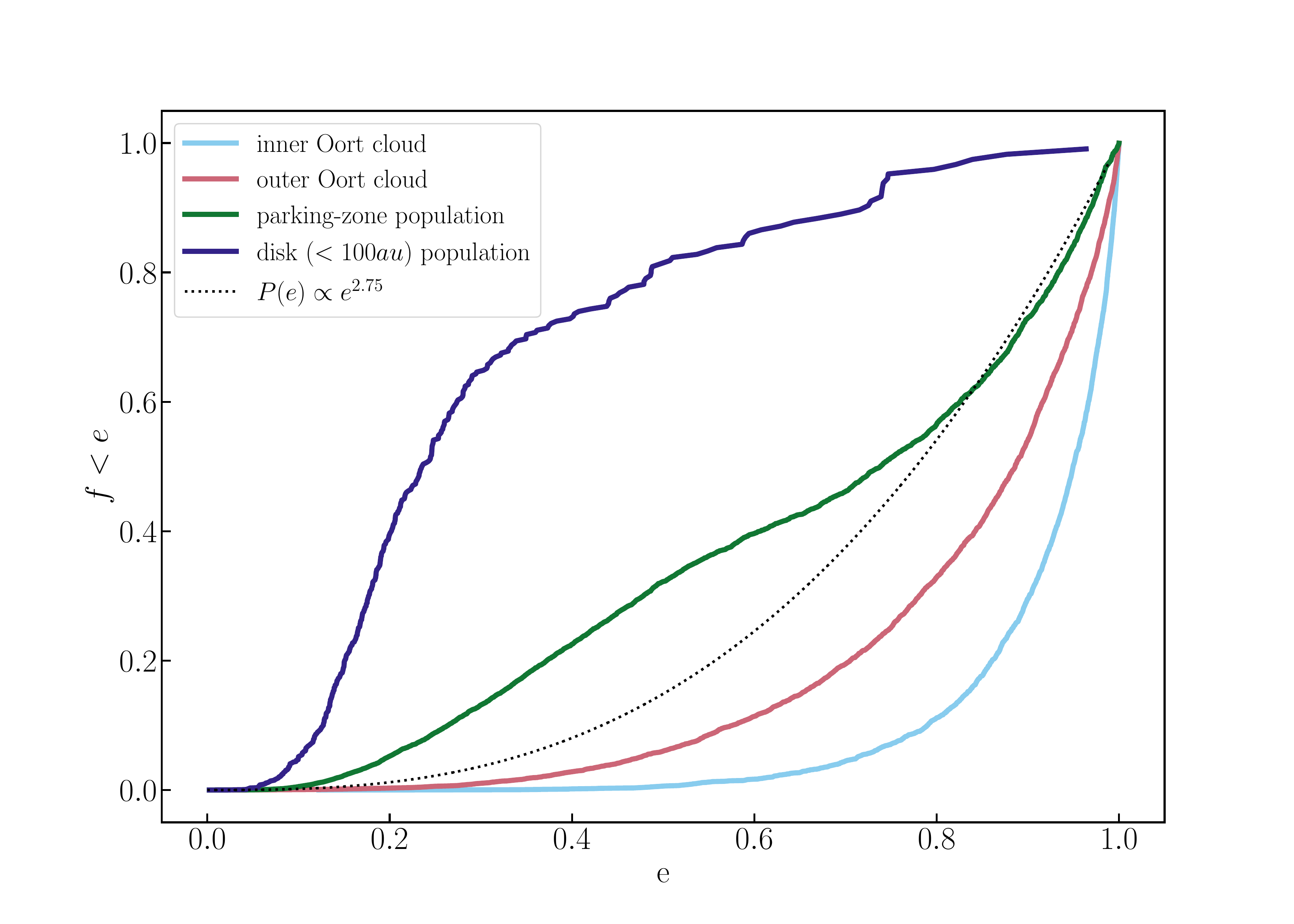}\\
\caption{Cumulative distribution of the eccentricity in the inner and
outer Oort cloud at an age of 50\,Myr after the start of the
simulations. The additional green curve gives the final distribution
of $P(e)\propto e^{2.75}$. The simulated eccentricity distribution
approach the theorectical curve in about 50\,Myr.}
\label{fig:eccentricity_distribution}
\end{figure}

\subsubsection{Integrating the equations of motion}\label{Sect:IC_D}

From the computed starting locations of the stars (see
figure\,\ref{fig:solar_orbit_in_the_galaxy_at_0Gyr}) and after planets
and asteroids are situated around each (see
figure\,\ref{fig:OCorbital_elements_init}), we integrate their
positions (and those of the minor bodies) forwards in time for
1.1\,Gyr.  Asteroids, in this calculation feel the gravity of the
star, the planets and the background potential of the Galaxy. The
planets around a specific star feels their sibling planets, the parent
star and the background potential. Each star feels its planets and the
background galactic potential.

After 1\,Gyr of integration all stars have arrived at (or at least
near) the currently observed Gaia DR2 position. In
figure\,\ref{fig:galactic_orbit_of_specific_stars} we show the
back-and forwards calculation for two stars in our sample, Glise 143
and 61 Cygnus A. For the former, the back and forwards calculations,
both remain within a line-width in the figure.

The yo-yo approach in a time-resolved smooth potential will result in
precisely the same forwardly and backwardly orbital trajectory, as can
be seen in the left panel of
figure\,\ref{fig:galactic_orbit_of_specific_stars}, where we
demonstrate this for Glise\, 143. However, the forward calculations do
not always precisely match the backward calculations because sometimes
one or more planets are ejected while integrating the stellar orbit
forwards in time. The result of a planet-ejecting system can be seen
in the right panel of
figure\,\ref{fig:galactic_orbit_of_specific_stars}, where we
integrated 61 Cyg A in the yo-yo approach. The forwards and backward
calculation, in that case, result in slightly different orbits.

The ejection of a planet causes the center-of-mass of the star and
planets to arrive at the anticipated Gaia DR2 location, but the
forwardly calculated stellar trajectory will deviate from the backward
calculation due to conservation of energy and linear momentum (in
section\,\ref{Sect:discussion} we discussion the consequences).

61 Cyg A is one of the worst cases in terms of the yo-yo approach,
because two planets of $1.79 M_{\rm Jupiter}$ and $0.72\,M_{\rm
  Jupiter}$ were ejected $\sim 105$\,Myr after the start of the
forward calculation.  The massive planet is ejected first with a low
relative velocity, but the lower-mass planet is ejected much more
violently. This second planet is visible in
figure\,\ref{fig:galactic_orbit_of_specific_stars}, to the far right
at a distance of about 7500\,pc from the Galactic center (see also
figure\,\ref{fig:solar_orbit_in_the_galaxy_at_1Gyr}).

\section{Results}

In the yo-yo approach, each star reaches its current location after
first being calculated backward in time and subsequently forwards
(with minor bodies). The forward calculation is performed with planets
and asteroids up to the current epoch and 100\,Myr 'back to the
future'. The current positions of the stars, planets and asteroids are
presented in
figure\,\ref{fig:solar_orbit_in_the_galaxy_at_1Gyr}. During the
calculations each planetary system is treated as isolated with respect
to the other planetary systems.

Our calculations start with a star with planets and a conveyor belt
filled with asteroids. The distributions in semi-major axis and
eccentricity of planets and asteroids are presented in
figure\,\ref{fig:OCorbital_elements_init}. While integrating forwards
in time, the conveyor belt is emptied and the asteroids either remain
bound or become unbound. In the next two sections, we discuss both,
planets and asteroids bound to the parent star or unbound.

\subsection{The population of bound planets and asteroids}

We recognize several regions where an object can have a bound orbit
around its parent star; it can be parked in the Kuiper belt or
deposited in an Oort cloud. No asteroids are trapped in a resonant
orbit with a giant planet.

\subsubsection{Asteroids in the Kuiper belt and the Parking zone}

In isolated planetary systems, as the ones simulated, the parking zone
is expected to be deprived of asteroids. Still, $0.32\pm0.01$\,\% of
the asteroids in the simulations are deposited in their respective
stars' Parking zone.  About one-third of these populate the area
within 100\,au along the conveyor belt's outer edge. This population
is visible in figure\,\ref{fig:OCorbital_elements_today} as the red
bullets between the right-most black solid curve and the right-most
cyan curve. There is even an odd planet in the parking zone.

These parked asteroids are driven there through several scatterings
with planets that have also migrated. Asteroids in the parking zone
did not migrate along the conveyor-belt but were scattered several
repeatedly by planets that themselves were scattered. As a
consequence, the asteroid did not move along the conveyor belt but
experiences a random walk until parked in a relatively wide ($100 < a
< 10^4$\,au) orbit with a relatively low eccentricity ($e\aplt 0.8$).
The right-most cyan curve in
figure\,\ref{fig:OCorbital_elements_today} indicates the outer edge of
the parking zone~\citep{2015MNRAS.451..144P} and the inner edge of the
Oort cloud.

\subsubsection{The formation and early evolution of the Oort cloud}

Once an asteroid passes the right-most cyan curve in
figure\,\ref{fig:OCorbital_elements_today} we consider it member of
the Oort cloud.

In figure\,\ref{fig:Oortcloud_fraction} we present the evolution of
the fraction of asteroids in the Oort cloud of their respective stars.
In 25\,Myr half the asteroids are removed from the planetary
system. At the age of 100\,Myr the fraction of asteroids in the disk
dropped to $\sim 2$\%, and after 1\,Gyr, only 0.3\,\% of the asteroids
still orbit in a disk within $\sim 100$\,au around the parent
star. All bound asteroids are eventually parked in some Kuiper
belt, in the Parking zone or in the Oort cloud.

The Oort cloud builts up on a time scale of a few 100\,Myr (see
figure\,\ref{fig:RHill_in_Galaxy_with_OC}). At the age of 10\,Myr, 4\%
to 6\% of the original disk population has been launched into the Oort
cloud. By that time, most orbits are still in the orbital plane of the
planets.

In figure\,\ref{fig:OCorbital_elements_today} we present the orbital
distribution of asteroids that remain bound to their respective stars.
The solid green curve indicates where the orbital period, $P_{\rm
  orb}$, equals the circularization-diffusion time-scale by the
Galactic tidal field ($t_{\rm diff}$, Eq.5 of
\cite{1987AJ.....94.1330D}). This line indicates the orbital
parameters for which a particle tends to spend more time being
perturbed by the Galactic tidal field (near apocenter) than by the
Sun. The orbits of asteroids (and planets to this curve's right, are
circularized and isotropized by the Galactic tidal field.  Asteroids
between the right-most cyan and the green curves have difficulty being
circularized by the Galactic tidal field; they tend to persist in
rather high eccentricity orbits. However, if an asteroid is pushed
further along the conveyor belt, deep into the Oort cloud, the
Galactic tidal field's circularization process operates on a time
scale shorter than the time scale on which the planets eject
asteroids. Asteroids that penetrate the Oort cloud further (to even
wider but bount orbits) tend to circularize more quickly. The mean
eccentricity in the Oort cloud is then a function of distance to the
parent star and of time.

In figure\,\ref{fig:eccentricity_distribution} we present the
cumulative distribution of the minor body's eccentricities in the Oort
cloud, 50\,Myr after the start of the simulation. Here we make the
distinction between an inner and an outer Oort cloud. This distinction
is made at a perturbation factor of $\delta v = 10^{-4}$. This
corresponds to a semi-major axis of about $6\times 10^4$\,au (for a
circular orbit), compared to the division between the outer edge of
the parking zone and the inner edge of the Oort cloud, which is at
$\delta v = 10^{-5}$ and corresponds to a semi-major axis of $\sim
10^4$\,au (right-most cyan curve in
figure\,\ref{fig:OCorbital_elements_today}).

\begin{figure}
\centering
\includegraphics[width=\columnwidth]{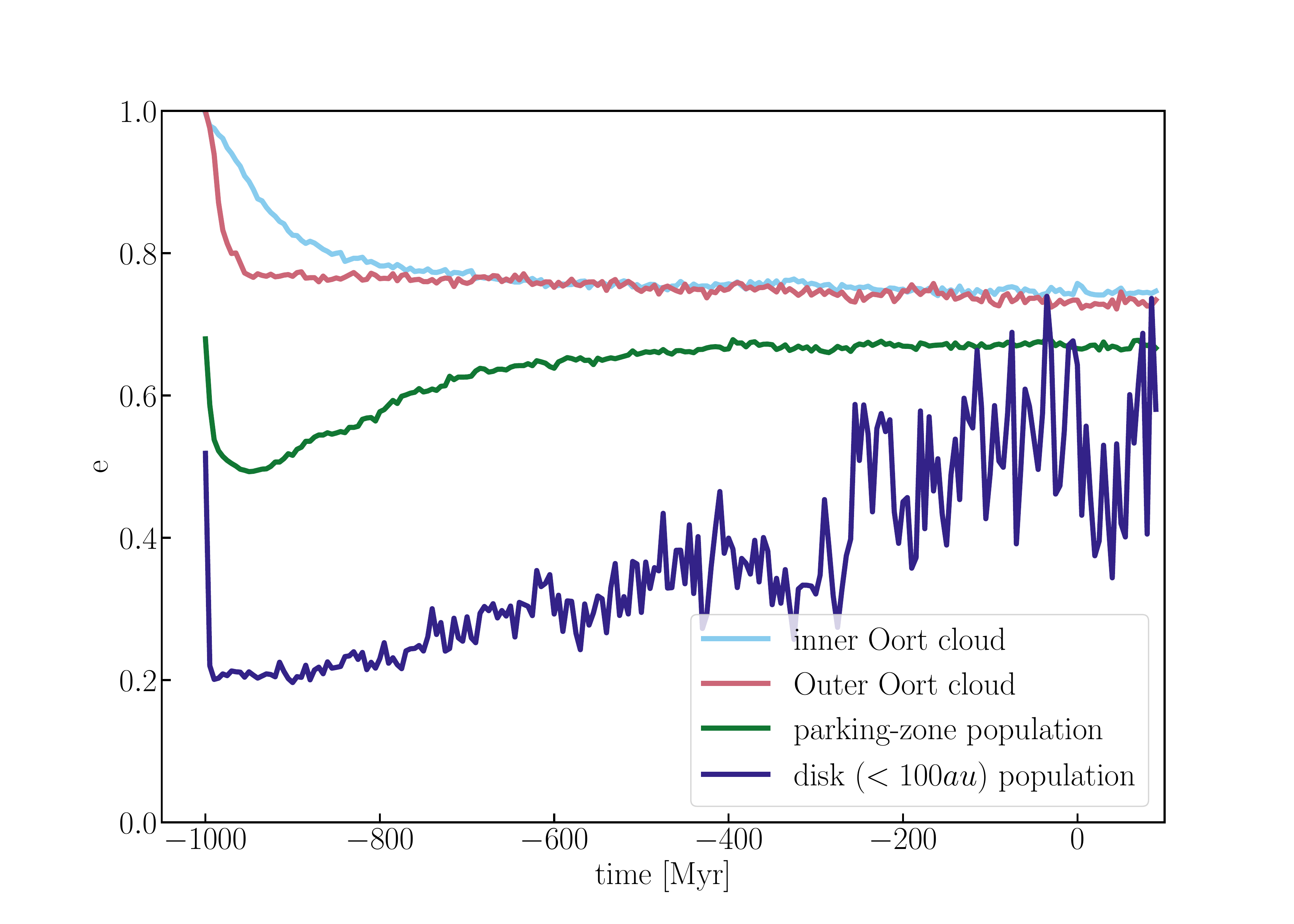}\\
\caption{Mean eccentricity as a function of time for the Oort cloud
  (inner and outer) the asteroids in the parking zone and the disk
  (within 100\,au). Initially, all asteroids have reativel high
  eccentricity, but this rapidly drops in the first few tens of Myr
  due to ejectsion by the planets. The Oort cloud population remains
  steady after about 100\,Myr, whereas the more inner regions slowly
  converge to a similar eccentricity of $\sim 0.7$.}
\label{fig:mean_eccentricity_evolution}
\end{figure}

In the first few tens of millions of years, when the Oort cloud is
populated, the inner and outer Oort clouds have quite distinct
distributions in eccentricity. This can be seen in the evolution of
the mean eccentricity in these regions, presented in
figure\,\ref{fig:mean_eccentricity_evolution}.

\subsubsection{The evolution of the isotropic Oort cloud}

After about 100\,Myr both eccentricity distributions approach the
thermal distribution. The Parking-zone population at this time is
still converging to the thermal distribution.  For a more thorough
discussion on the asteroid migration in the Solar system, we refer to
paper I (Portegies Zwart et al.\, in preparation).

Once in the Oort cloud, the orbital distribution of asteroids becomes
isotropic in inclination and their eccentricity drops from $\gg 0.9$
to a mean eccentricity of $\langle e \rangle \sim 0.7$ (see
figure\,\ref{fig:mean_eccentricity_evolution}).  200\,Myr after the
start of the simulation, the Oort clouds are fully developed (see also
figure\,\ref{fig:RHill_in_Galaxy_with_OC}).

After reaching a maximum, the number of asteroids in the Oort cloud
starts to decay on a half-life time-scale of about 800\,Myr. This time
scale is considerably shorter than estimates for the Sun's Oort cloud
derived by \cite{2018MNRAS.473.5432H}.  Our calculations accounted for
the variations in the smooth Galactic potential through which the star
moves, and part of the discrepancy stems from the wide range in the
ellipticities of the Galactic orbits of nearby stars (see
figures\,\ref{fig:galactic_orbit_of_specific_stars} and
\ref{fig:RHill_in_Galaxy_with_OC}).  The orbit of the Sun in the
Galactic center, was also reason for \cite{2011Icar..215..491K} to
discuss variations in the inner edge of the Oort cloud around the Sun.
Our calculations \citep[and those of][]{2011Icar..215..491K} neglected
the effect of passing stars, which tends to dramatically reduce the
Oort cloud's lifetime
\citep{1986gss..conf..204W,1990Icar...84..447S,1992CeMDA..54...13M,2018MNRAS.473.5432H}.

\subsubsection{The population of bound planets}

Planets experience a similar fate as the asteroids, but it is much
harder to eject a massive planet than a mass-less asteroid. Therefore,
most planets remain bound to the parent star, and stay within the
planetary region originally between 10 and 100\,au. In
figure\,\ref{fig:OCorbital_elements_today} it is shown how the planets
scatter along the curves of constant pericenter (flaring to the right)
and constant apocenter (flaring to the left).

About $40$\,\% of the planets are ejected in the first
200\,Myr. Eventually, after 1\,Gyr, $\sim 24.8$\,\% of the planets
remain bound to their respective stars, and $\sim 71.6$\,\% is ejected
to become free-floating. Only $\sim 2$\,\% of the planets acquire a
bound orbit in their star's Oort cloud. The orbital distribution of
these is isotropic in inclination with relatively low eccentricity
($\langle e \rangle \sim 0.7$, see
figure\,\ref{fig:mean_eccentricity_evolution}).  In
figure\,\ref{fig:OCorbital_elements_today}, we present the orbital
distribution of planets and asteroids that remain bound to their
respective stars.

Planets in an Oort cloud are rare. In all our simulations, ten out of
$\sim 736$ planets are parked in the Oort cloud. Only $\sim 5$ percent
of the stars in our simulated solar neighborhood have a planet in the
Oort cloud.

\subsection{The population of unbound planets and asteroids}

Most asteroids and planets do not remain bound to their host
star. These objects become rogue planets and asteroids.

\subsubsection{Rogue asteroids and \soli}

\begin{figure}
\centering
\includegraphics[width=\columnwidth]{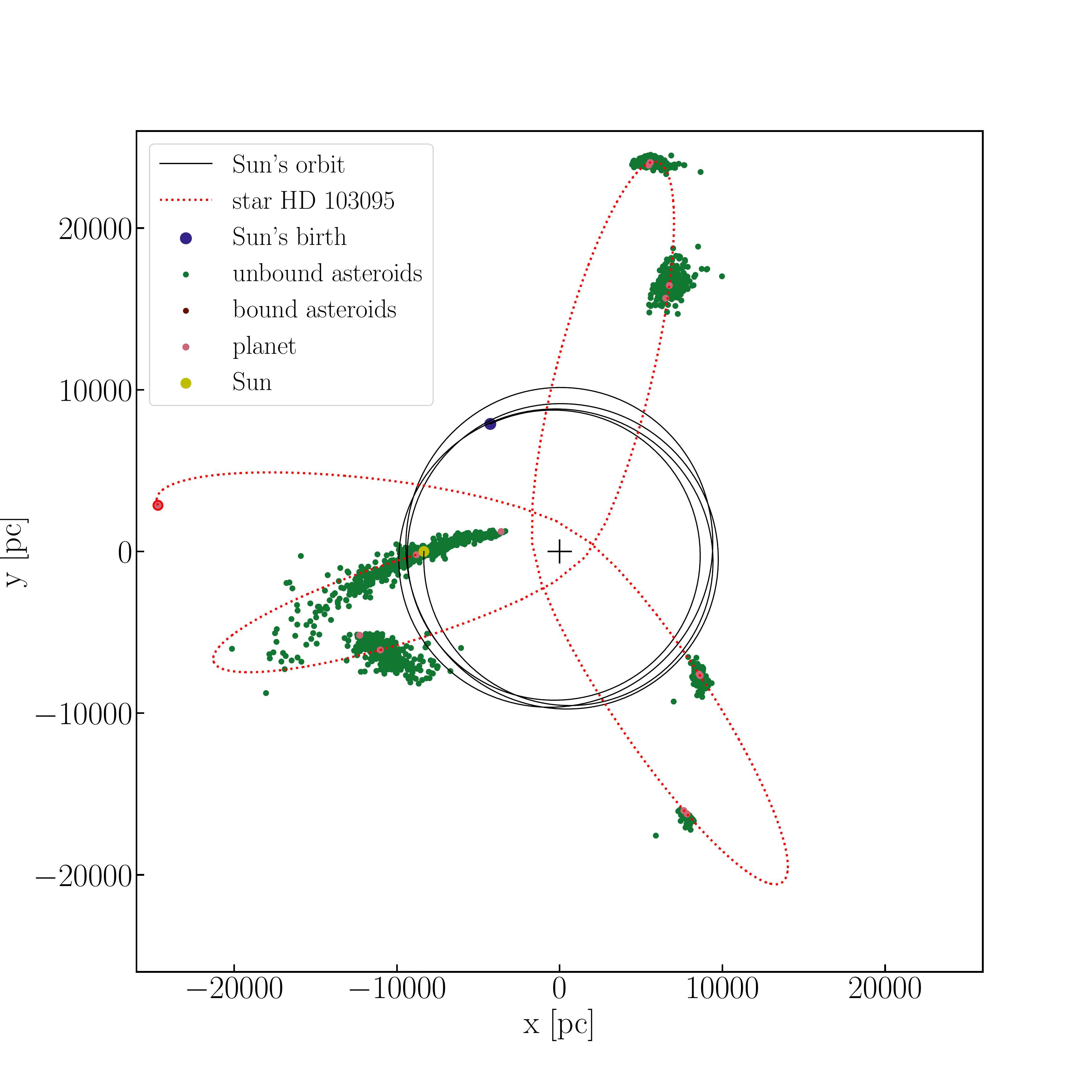}
\caption{Projected Galactic orbit for HD 103095, which lost all its
asteroids after the first passage of the Galactic center. HD 103095
(Groombridge 1830) is a is a metal-poor star ($\log(g) = 4.71 \pm
0.3$ and $[Fe/H] = -1.27\pm0.2$\,dex) \cite{2018MNRAS.475L..81K})
with a high proper-motion. According to Gaia DR2 its mass is $\sim
0.87$\,\MSun\, (but only 0.66\,\MSun\, according to
\cite{2007ApJS..168..297T}). In our calculations this star has a
rather elongated orbit in the Galactic potential. Further
description as in the legend of
figure\,\ref{fig:galactic_orbit_of_specific_stars}.}
\label{fig:galactic_orbit_of_specific_stars_HD}
\end{figure}

The vast majority of asteroids become free-floating in the potential
of the Galaxy. In figure\,\ref{fig:galactic_orbit_of_specific_stars}
we present the evolution of the unbound asteroids for Glise 143 and 61
Cyg, an for HD 103095 in
figure\,\ref{fig:galactic_orbit_of_specific_stars_HD} along 5 (equally
spaced) moments in time. At the start of the simulation, all asteroids
(and planets) are bound to the star, but after 200\,Myr most asteroids
have already become unbound.

Although the fraction of bound asteroids is generally small (see
figure\,\ref{fig:Oortcloud_fraction}), there are always some asteroids
in the Oort cloud. Except for the halo star HD 103095 (Gaia DR2
4034171629042489088), which has lost all its asteroids at 100\,Myr
after the start of the simulation; No Oort cloud was formed around
this star. The close perigalactic distance of this star together with
the chaotic reorganization of its planets, resulting in the ejection
of the lowest-mass planet ($\sim 1.59\,M_{\rm Jupiter}$), made it hard
to preserve any objects bound to its Oort cloud. Its projected orbital
trajectory is depicted in
figure\,\ref{fig:galactic_orbit_of_specific_stars_HD}.

Escaping asteroid tends to hang around the parent stars \citep[see
  also][who performed similar calculations for the Solar
  system]{2019MNRAS.490.2495C}. Much in the same way as tidal tails
from star clusters tend to hang around along the trailing and leading
orbital trajectory \citep{2017MNRAS.470..522S}. A similar structure is
observed along the orbit of the Sagittarius dwarf
\citep{2003ApJ...599.1082M}, Pal 5 \citep{2001ApJ...548L.165O} and
other globular clusters \citep{2020A&A...637L...2P}. After the first
$\sim 200$\,Myr, there has been insufficient time, for the unbound
asteroids to drift far away from the parent star, and the tidal
asteroidal debris tails have a length of at most a few parsecs. After
600\,Myr the debris tails have fully developed, extending over several
kpc. The smearing and rotation of the debris tails is then determined
by their orbital phase on the Galactic potential. For most of the
orbit, the tidal tails closely follow the parent star's orbit.

\subsubsection{Rogue planets}

The planetary systems in our simulations tend to be born in moderate
stability, and some planets are ejected while interacting with each
other or with the tidal field of the Galaxy.  But this planet-ejection
process is less common because of their large mass compared to the
asteroids.  Part of these planet ejections can be the result of
numerical errors, because these systems are highy chaotic.  The
largest positive Lyapunov exponent indicates a stability time-scale
considerably shorter than the time scale over which we integrate these
systems. Although the adopted direct N-body code ({\tt Huayno}) for
integrating the planetary systems, there is some secular growth of the
energy error, which in combination with the coupling to the Galactic
potential may lead to spurious planet ejections.

A consequence of planet-planet scattering is the preferable ejection
of lower mass planet, leaving the more massive planets in perturbed
(possibly inclined and eccentric) orbits
\citep{2018MNRAS.474.5114C}. This is illustrated in
figure\,\ref{fig:initial_planet_mass}, where we present the initial
mass-function of planets, the final mass-function (after 1\,Gyr), and
the mass function of the escaped planets. The final bound planet mass
is larger than the initial planet mass, due to the relatively common
escape of low-mass planets. Initially half the planets have a mass of
$\sim 2.5\,M_{\rm Jupiter}$, whereas the mean mass of escaping planets
is about $\sim 1\,M_{\rm Jupiter}$.

In our calculations, the effect of the Galactic tidal field on the
dynamical evolution of the planets is small, except for those stars
that have a small peri-galactic orbit (such as HD\,103095). The strong
perturbing effect of the Galactic center led
\cite{2013MNRAS.430..403V} to argue that the Galactic-center region
will have a higher proportion of rogue planets than the relative
outskirts of the Solar neighborhood.

\subsection{The current local Galactic distribution of asteroids}

In figure\,\ref{fig:solar_orbit_in_the_galaxy_at_1Gyr} we present the
distribution at the current epoch 1Gyr after the start of the
simulation. It is no coincidence that at this moment, most objects are
located around and near the Sun. Some structures along the Sun's orbit
in the Galactic potential and some streams perpendicular to the
current Sun's orbits are discernible. In the coming paragraphs, we
further explore this population by presenting various slices in
parameter space. The objective of this study is to better understand
the local phase-space distribution of \soli.

\subsubsection{Characterization of the local distribution of \soli}

\begin{figure}
\centering
\hspace*{-0.8cm}
\includegraphics[width=1.0\linewidth]{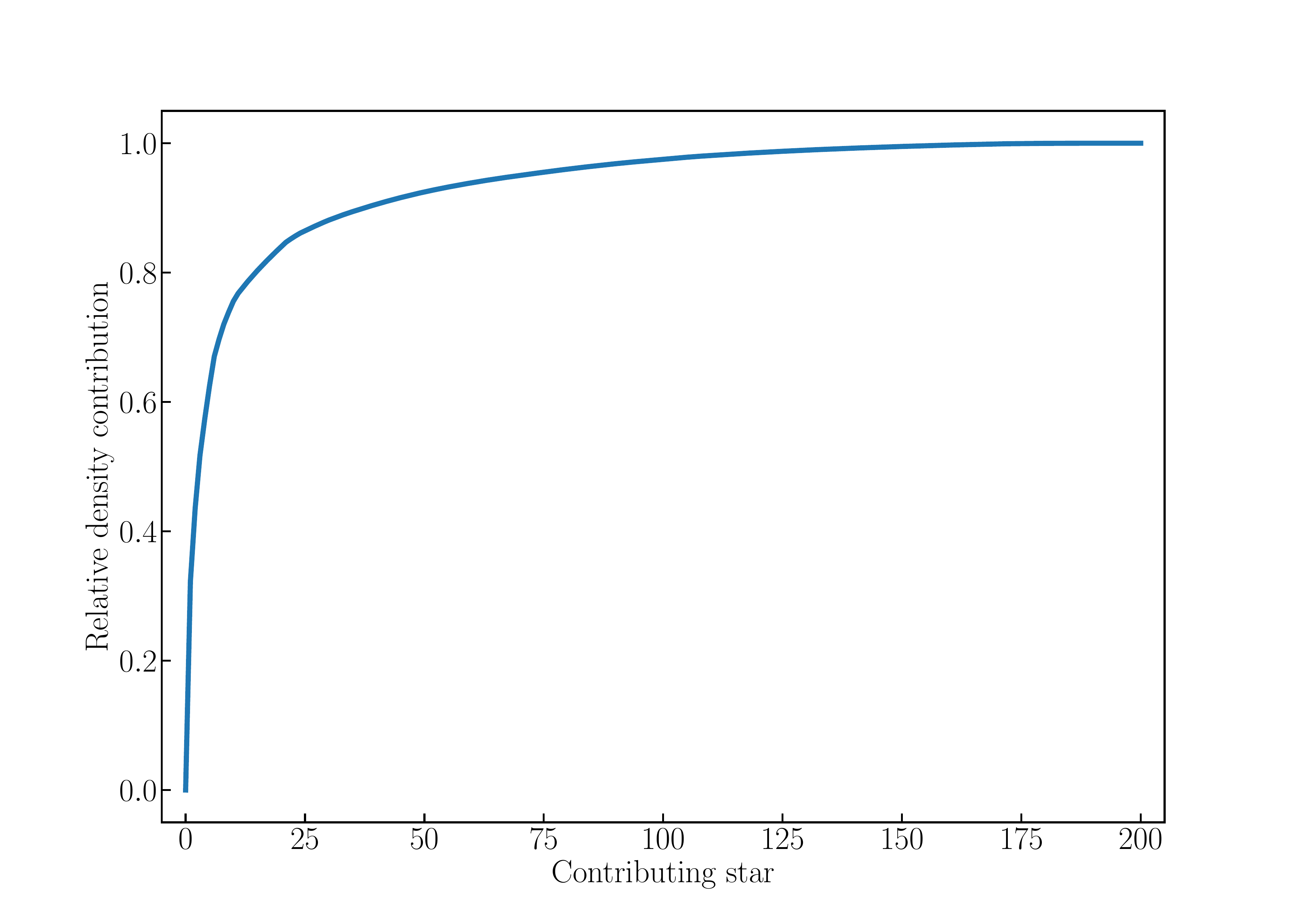}
\caption{Relative cumulative contribution to the local density of
minor bodies in the Solar neighborhood (sorted on their contribution
to the local density of \soli). We used a KDE smoothing kernel with
1.41\,pc.}
\label{fig:relative_density_contribution}
\end{figure}

\begin{figure}
\centering
\includegraphics[width=\linewidth]{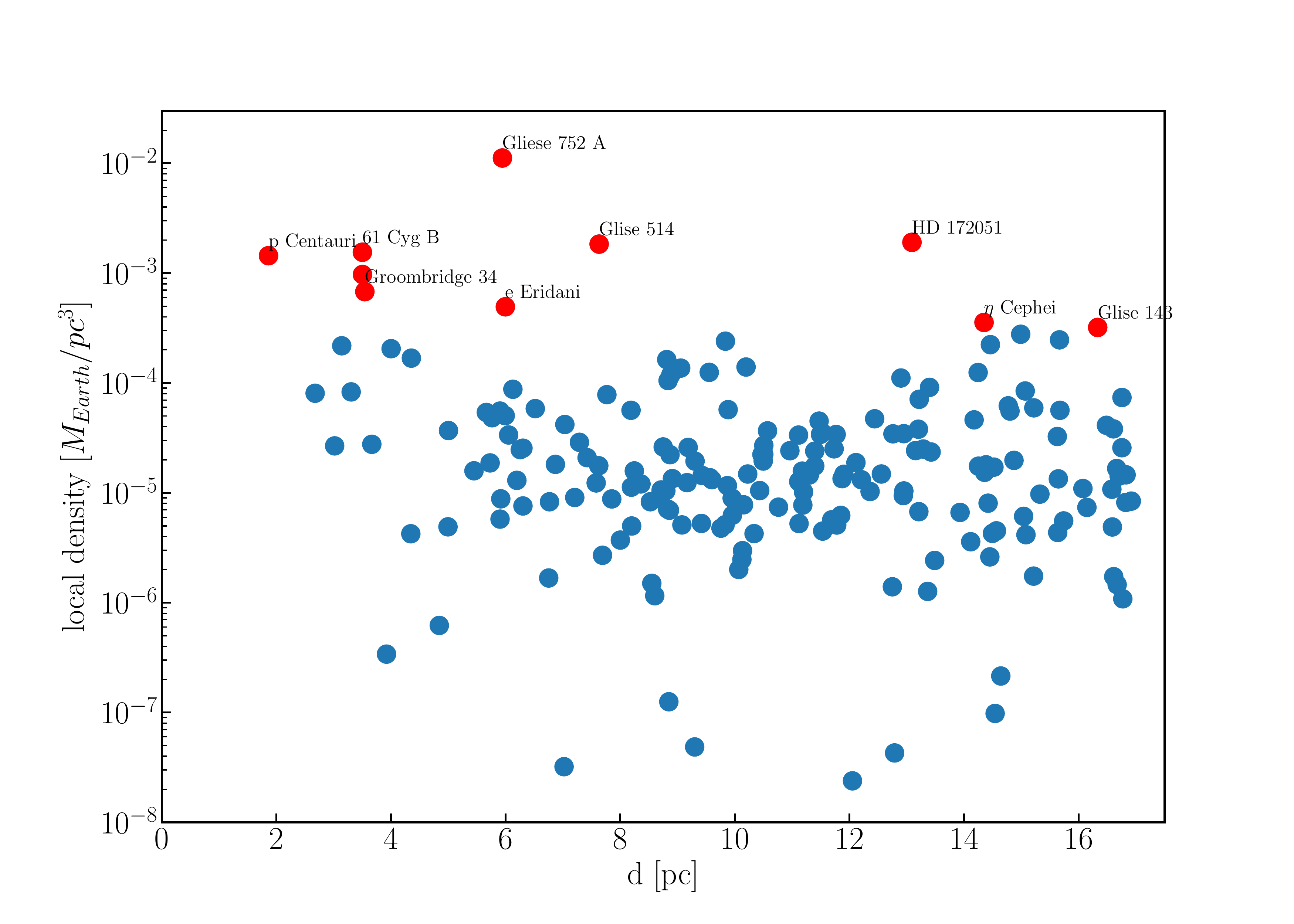}\\
\caption{Individual contribution to the local density of \soli\,
calculated using a kernel density estimator with a smooting length
of 0.39\,pc. The ten biggest contributors to the local density are
idicated in red and identified with their popular name. }
\label{fig:density_contribution_at_current_epoch}
\end{figure}

In figure\,\ref{fig:relative_density_contribution} we present the
relative cumulative contribution of the local density of interstellar
objects in the vicinity of the Solar system of the 200 nearby stars.
In figure\,\ref{fig:density_contribution_at_current_epoch} the
contribution for each individual star is sorted to its current (Gaia
DR2) distance. The 10 most contributing stars to the local density of
\soli\, are painted red, all the others are blue. 

The 10 most contributing stars are listed in
table\,\ref{tab:nearby_contributors}, and contribute more than 70\% to
the local density.  About half the stars hardly contribute to the
current local density of interstellar minor bodies. Even if the star
is close to the Solar system, the stream of unbound asteroids may
simply miss our current location in the Galaxy, as is the case for the
second to fifth nearest stars to the Sun (see
table\,\ref{tab:nearby_contributors}), which contribute less than
$\eta$~Cephei or Glise 143.

The total local density of \soli\, $\sim 0.027\,M_{\rm
  Earth}/$pc$^3$. With an estimated mass for 'Oumuamua of $\sim 1.2
\times 10^{9}$\,kg \citep{2018MNRAS.479L..17P}, the nearby stars
contribute to the local density of interstellar asteroids by $\sim
1.2\times 10^{14}$ per cubic parsec.

The local density of interstellar minor bodies was estimated by
\cite{2018MNRAS.479L..17P,2018ApJ...855L..10D} to be of the order of
$10^{15}$ objects per pc$^{-3}$. This density has contributions from
evaporating Oort clouds or from the disruption of circumstellar debris
disks. Both tends to be similar in scope in our analysis, because Oort
cloud formation tend to be associated with debris-disk destruction
\cite[although more gently than the process discussed in ][who adopted
  that the disk would be obliterated by the post-asymptotic
  giant-branch evolution of the parent
  star]{2011MNRAS.417.2104V,2013MNRAS.430..403V}.

It is hard to estimate how much stars further away contribute to the
local density of \soli, in particular since we just demonstrated that
far-away stars may contribute considerably when the Sun happens to
move through the tidal-debris tail.  We then estimate that Oort cloud
evaporation and debris-disk distruction contribute $\apgt 10$\,\% to
the observed population of \soli.


To further illustrate this, the list in
table\,\ref{tab:nearby_contributors} is sorted to the distance to the
sun (n-dist $=1$ for the closest star p Centauri). Some nearby stars
contribute considerably to the local density, but many major
contributors are further away, as far as Glise 143 at about
16\,pc. Note that the contribution to the local density depends
somewhat on the adopted smoothing kernel for the density
estimator. When we adopted a smoothing kernel of 1.0\,pc instead of
0.38\,pc, the star SZ UMa (the 70th nearest to the Sun) contributes
almost as much to the local density of \soli\, as 61 Cyg~B.

\begin{table*}
\centering
\caption{List of the 9 nearby stars that contribute most to the local
(at the Solar system today) density of \soli. Listed are the Gaia
DR2 identifier, the common name, the n-th closest star, the actual
distance (in parsec) and velocity (in km/s) from the
Solar system. In the 6th and 7th columns, we present the
contribution to the local density of \soli\, in number per pc$^{-3}$
and as a relative fraction. The last two columns give the mass of
the star according to the most reliable estimates, according to the Gaia data archive,
}
\begin{tabular} {llrrrrrrr}
\hline
Gaia DR2 id & name & n-dist & dist & v-rel& \multicolumn{2}{|c|}{density} & Gaia mass & mass \\
\hline
& & & [pc] & [km/s] & [pc$^{-3}$] &relative& \multicolumn{2}{c}{[\MSun]} \\
\hline
\noalign{\smallskip}
4293318823182081408 & Glise~752~A & 23 & 5.94 & 54.1 & 0.0111 & 0.42 & 0.63 & 0.46 \\
4079684229322231040 & HD 172051 & 135& 13.09 & 36.1 & 0.0019 & 0.07 & 1.07 & 1.0 \\
3738099879558957952 & Glise 514 & 44 & 7.63 & 57.9 & 0.0018 & 0.07 & 0.61 & 0.54 \\
1872046574983497216 & 61~Gyc~B & 6 & 3.50 & 107.3 & 0.0015 & 0.06 & 0.64 & 0.63 \\
4472832130942575872 & p Centauri & 1 & 1.86 & 142.6 & 0.0014 & 0.05 & 0.49 & 0.144 \\
1872046574983507456 & 61~Gyc~A & 7 & 3.50 & 109.0 & 0.0010 & 0.04 & 0.66 & 0.70 \\ 
385334230892516480 & Groombridge 34 & 8 & 3.54 & 50.7 & 0.0008 & 0.03 & 0.58 & 0.38+0.15 \\
4847957293277762560 & $\epsilon$~Eridani & 26& 6.00 & 124.9 & 0.0005 & 0.02 & 1.01 & 0.85 \\
2195115561163064960 & $\eta$ Cephei & 154& 14.35 & 105.0 & 0.0004 & 0.01 & 0.74 & 1.6\\
4673947174316727040 & Glise 143 & 185& 16.33 & 68.0 & 0.0003 & 0.01 & 0.72 & ? \\
\hline
\noalign{\smallskip}
\hline
\end{tabular}
\label{tab:nearby_contributors} 
\end{table*} 

\begin{figure}
\centering
\includegraphics[width=\columnwidth]{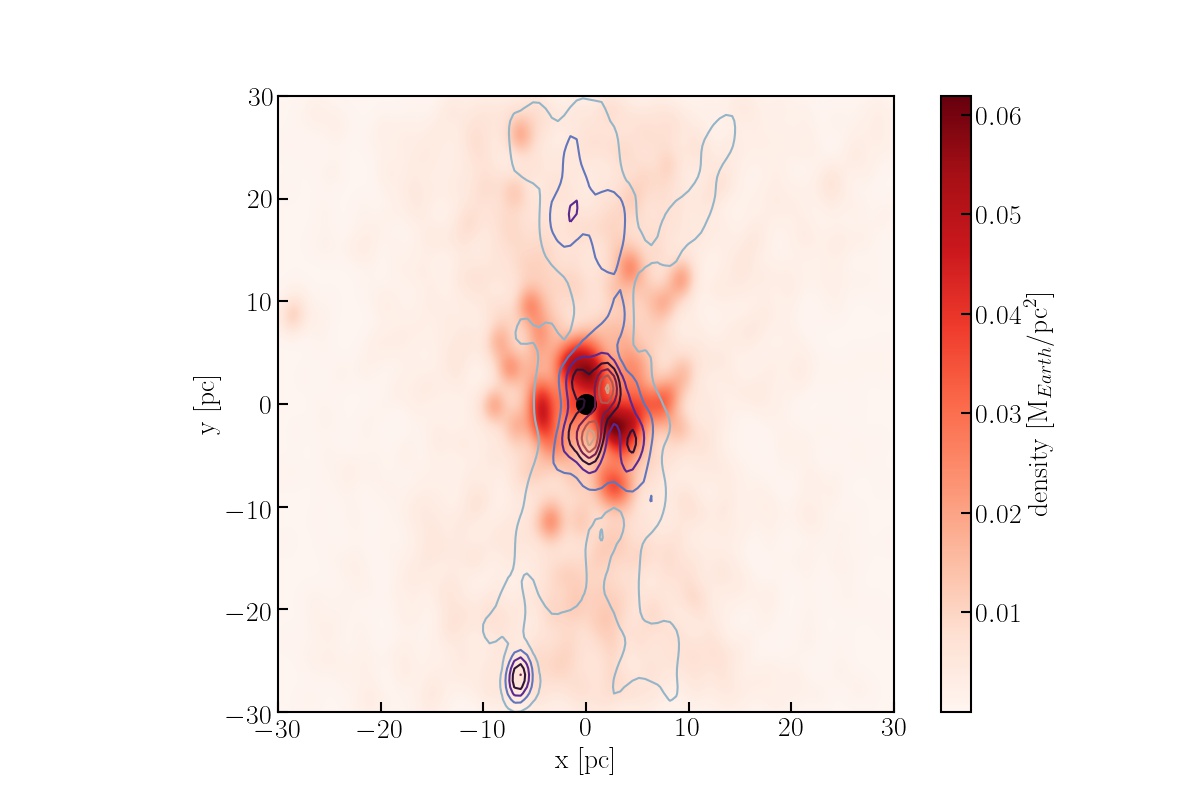}
\includegraphics[width=\columnwidth]{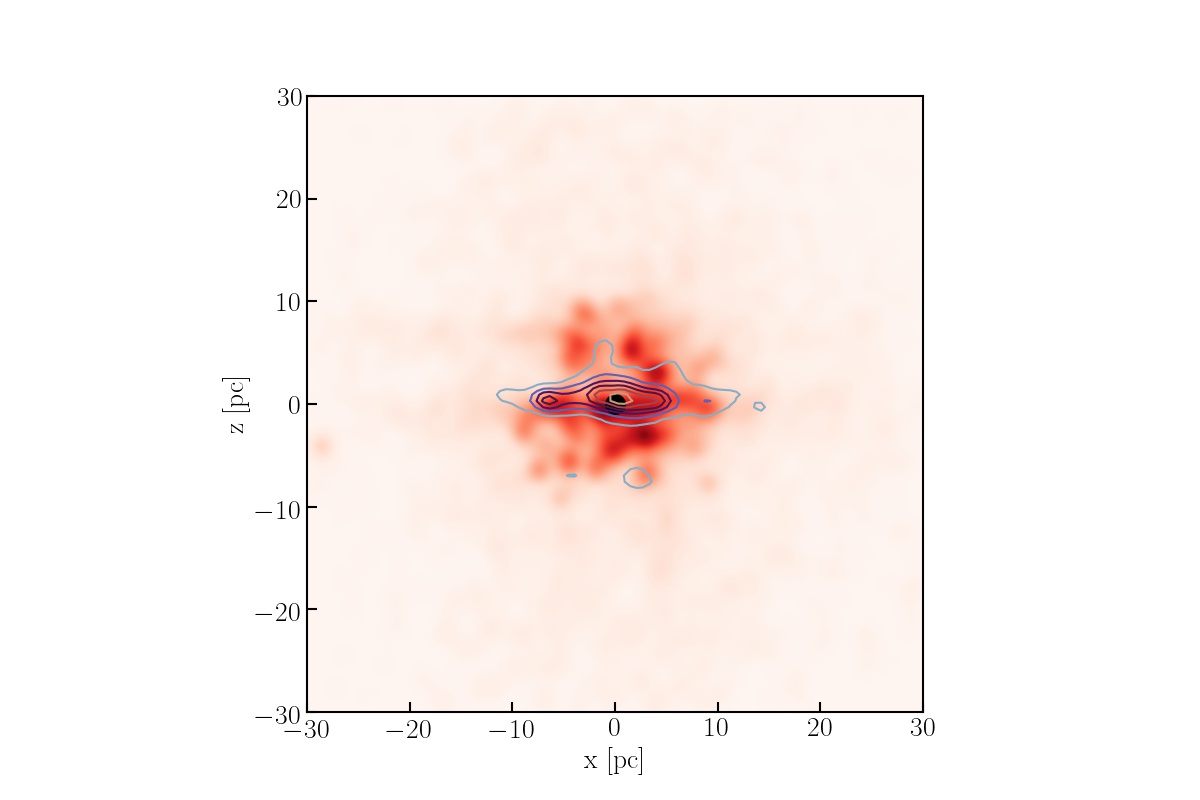}
\caption{Local projected density of asteroids in the solar
  neighborhood (within 30\,pc) in two projections: on the $X$-$Y$
  plane for the upper panel is a projection in the galactic plane,
  whereas the bottom panel shows the $X$-$Z$
  projection. Kernel-density smoothing kernel used was 0.104\,pc.  The
  red shades give the projected density of asteroids around the
  Sun. The contours give the local density only for the asteroids that
  originally orbited one of the ten stars in
  table\,\ref{tab:nearby_contributors}.}
\label{fig:local_density_xy}
\end{figure}

In figure\,\ref{fig:local_density_xy}, we present the stars that
contribute most prominently to the local minor-body density (see
table\,\ref{tab:nearby_contributors}) are presented as contours,
whereas the others are indicated with a color-shaded kernel-density
estimator. This distribution is clumpy, due to the minor bodies that
escaped the gravitational pull of their parent stars with a relatively
low velocity.

\begin{figure}
\centering
\includegraphics[width=\columnwidth]{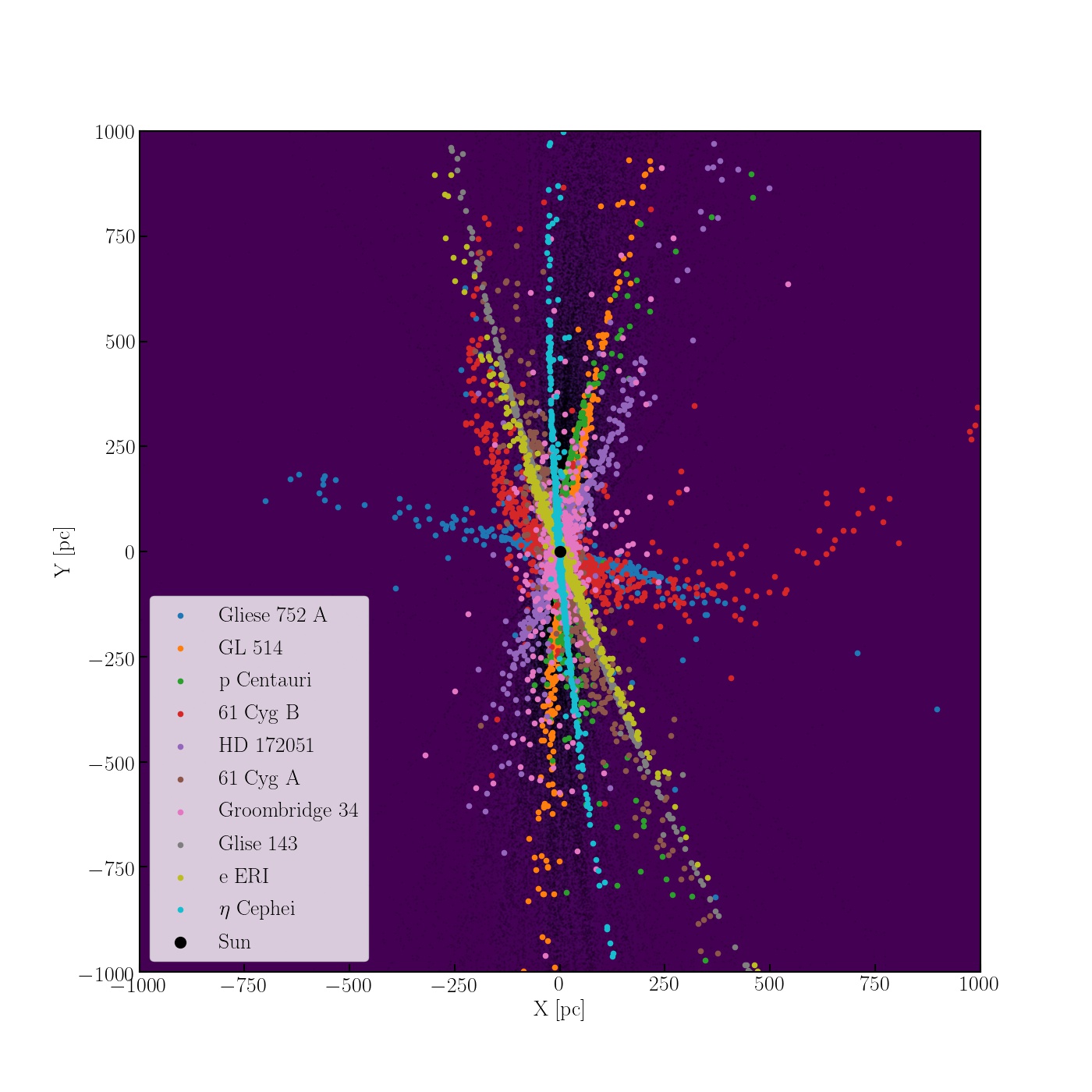}
\caption{Current density of asteroids within 1\,kpc of the Sun in the
Galactic xy-plane and centered around the Sun (black bullet in the
middle). The Galactic center is to the right. The 10 stars that
contribute most to the local density of \soli\, are indicated in
colors (see legend), the others as small gray dots. }
\label{fig:projection_nearby_stars}
\end{figure}

The debris streams are hardly visible in
figure\,\ref{fig:local_density_xy}, because they tend to extend over a
longer range and the relatively high local density causes the
subtleties of the extended tails to wash away in the background.
However, when we present the \soli\, of the most contributing stars
from table\,\ref{tab:nearby_contributors} on a larger scale the tidal
debris streams become evident. This is illustrated in
figure\,\ref{fig:projection_nearby_stars}. Here the colored bullets
identify the minor bodies of the 10 most important contributors. The
Sun is in the origin because this is how the contributing stars were
selected as nearest neighbors to the Sun (on which the picture was
centered). The morphology and width of the streams depend on the orbit
of the host star and orbital phase (see for example
figure\,\ref{fig:galactic_orbit_of_specific_stars}). A rather circular
orbit around the Galactic center results in a narrow and elongated
stream, as for the star $\epsilon$~Eridani, whereas a more elliptical orbit,
such as for Glise 143, results in a broader distribution along its
orbit. The interesting horseshoe shape of the tidal arms results from
Glise 143 being close to apocenter in its orbit around the Galactic
center: the trailing arm has the same spread as the leading arm but
being located closer to the Galactic center it overtakes the leading
arm.

\subsubsection{The closest approaching objects}\label{Sect:Close_approach}

The majority of the Solar neighborhood stars experience their
closest approach within a few hundred thousand years of the current
epoch (see also \cite{2019A&A...629A.139T}). The asteroids originally
belonging to these stars swarm in front of-- and behind them (as seen
in figure\,\ref{fig:solar_orbit_in_the_galaxy_at_1Gyr}). These tidal
tails may cause asteroids to pass the Sun millions of years before or
after the star has its closest approach to the Sun. We study
this phenomenon to understand the extent to which other stars
contribute to the local density of asteroids.

For this purpose, we analyze the same stars and minor bodies for the
last 20\,Myr and into the coming 20\,Myr in the Galactic potential
while keeping track of the closest approaches (the data is available
in the {\tt data/NN} directory in {\tt figshare}). In this time frame,
each asteroid has one closest encounter with the Sun.  \Soli\, tend to
arrive in families (as is also illustrated in
figure\,\ref{fig:projection_nearby_stars}). Each family originates
from a single parent star.  These families form from asteroids that
escape the stellar gravity well, and continue in a similar orbit
around the Galactic center as the parent star.  A particular star can
therefore lead to multiple close encounters with the Sun.  These
encounters may be spread out over the time-scale it takes the Sun to
cross the tidal-debris tail, and this may be several Myr.  During this
time frame the Sun may have several encounters with different
asteroids that originated from the same star. If several such
interstellar asteroids were found, it would be much easier to find the
source star.

In our simulations, the current dominant contributor to \soli\, seems
to be Glise 752.  In reality, Glise 752 may not contribute any \soli\,
simply because, contrary to our assumptions, it may not have an Oort
cloud.  Glise 752, is a known binary system with a possible planet
\citep{2018A&A...618A.115K}, but this does not make it necessarily a
candidate responsible for the processes described here.

\begin{figure*}
\centering
\includegraphics[width=\linewidth]{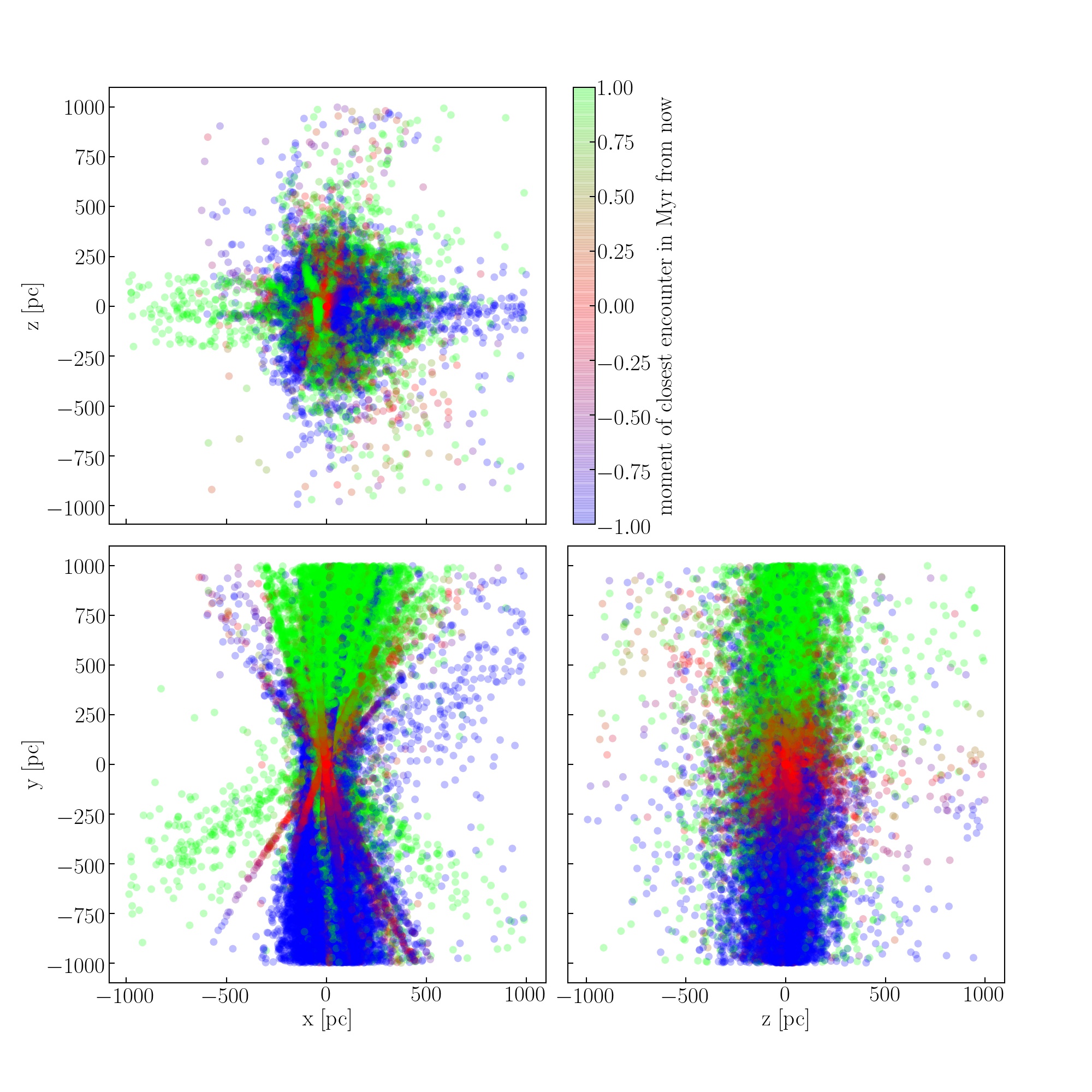}
\\
\caption{The moment of closest approach relative to the current
epoch along the three major axes. Asteroids in the wake of the
Solar system tend to have been encountered a while in the past,
whereas those in front of we will have the closest encounter in the
future.}
\label{fig:moment_of_closest_approach}
\end{figure*}

Most asteroids from the current neighboring stars tend to have their
closest approach with the Solar system in a rather short period of
only a few Myr. Even those further away, or the high proper-motion
halo-stars in the sample (see for example
figure\,\ref{fig:galactic_orbit_of_specific_stars_HD}), tend to have
their close approach within a few Myr of the current time. Since we
selected the 200 nearest stars, this is not a surprise; it reflects
how the current nearest stars were selected.

\begin{figure}
\centering
\begin{tabular}{c}
\includegraphics[width=\columnwidth]{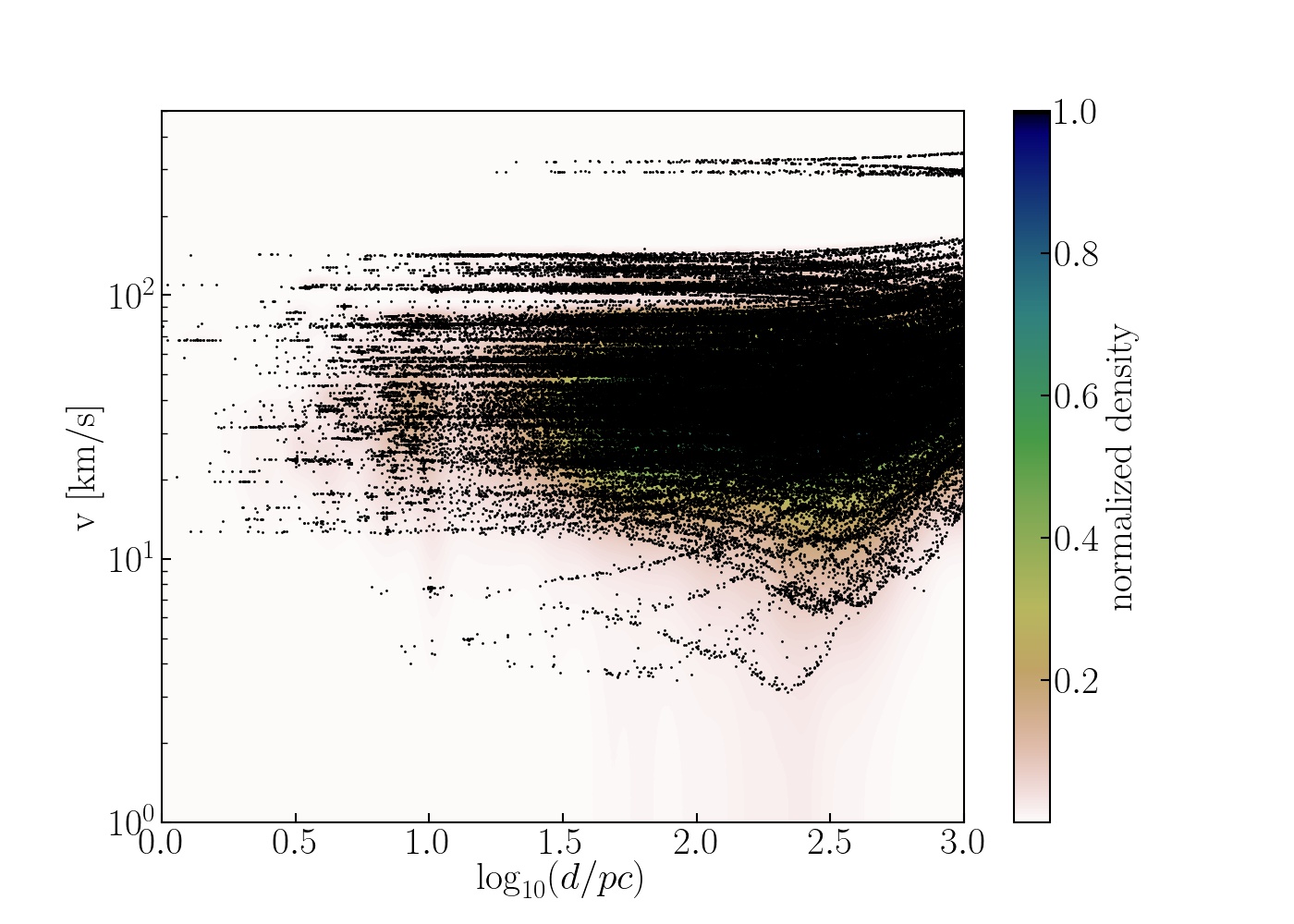} \\
\end{tabular} 
\caption{Relative velocity at closest approach versus the distance of
closest approach. The individual families of objects can be
identified as individual spikes to short distance. To guide the eye
we overplot the actual (unsmoothed) data with small dots. }
\label{fig:distvel_contours}
\end{figure}

The majority of stars orbit the Galactic center in the same general
direction as the Sun, but a few have retrogade orbits with respect to
the Sun. They are visible in
figure\,\ref{fig:moment_of_closest_approach} as the green bullets
among the blue bullets (and vice versa).  These retrogade orbits can
be identified in figure\,\ref{fig:distvel_contours} as those with a
large relative velocity with respect to the Sun.

\begin{figure}
\centering
\begin{tabular}{c}
\includegraphics[width=\columnwidth]{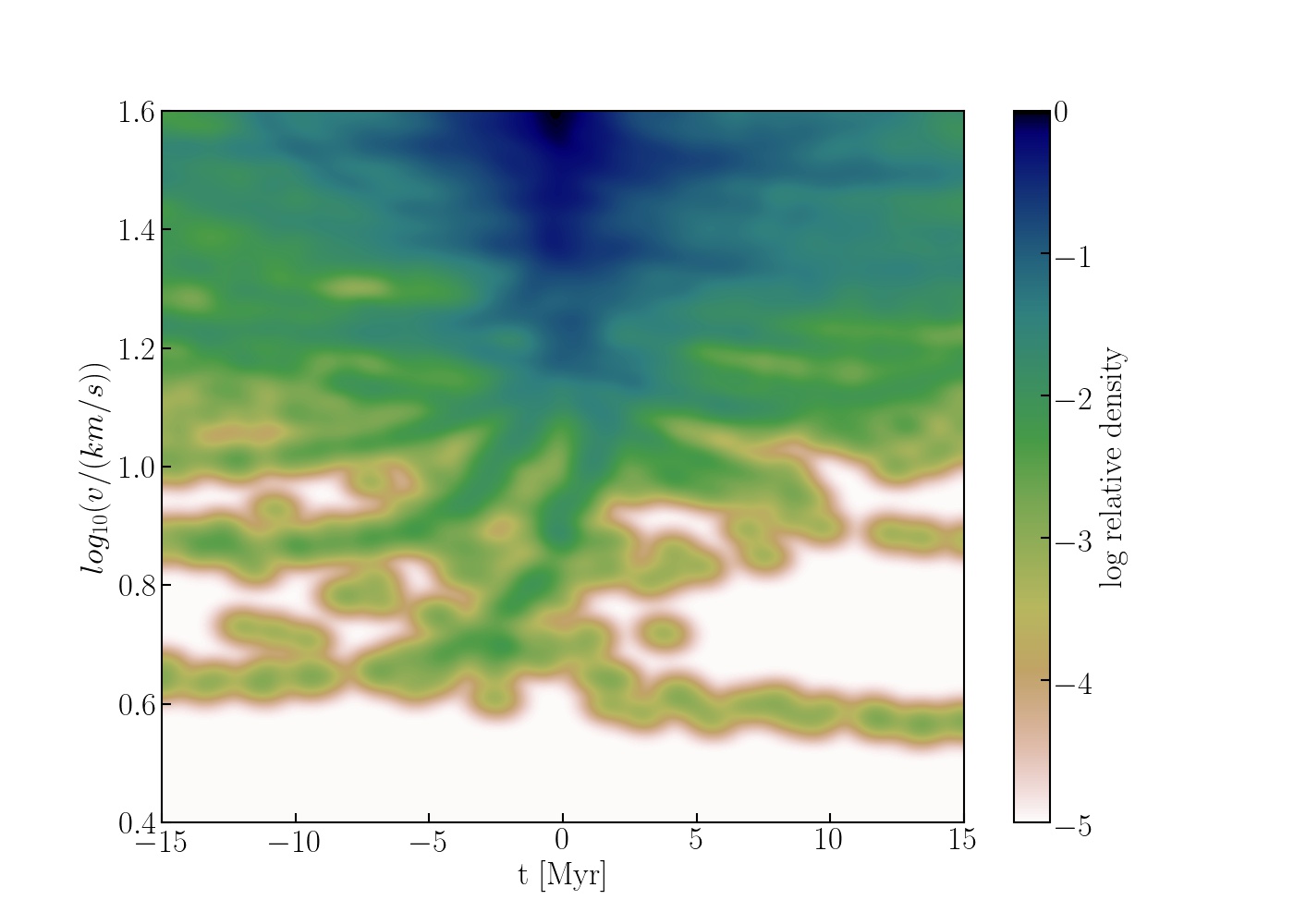}\\
\end{tabular} 
\caption{Distribution of relative velocity at the closest approach
with the Sun as a function of moment of closest approach. The color
colding gives density normalized to the maximum. The streams of
objects with coherent motion in time represent families of asteroids
belonging to a specific star.}
\label{fig:timevel_contours}
\end{figure}

The streams of asteroids in the left and right wings in
figure\,\ref{fig:moment_of_closest_approach}, are the result of
families of objects belonging to specific stars that pass today's
Solar system. Asteroids released around the stars' leading tidal tail
in the potential of the Galaxy have a close encounter with the Sun at
an earlier epoch, whereas asteroids that escape through the trailing
Lagrangian point have their closest approach to the Sun at a later
time. These streamers are even more pronounced in
figure\,\ref{fig:timevel_contours} where we present the relative
velocity at closest approach as a function of time of closest
approach. Specific families of asteroids are now clearly visible, in
particular at relatively low relative velocity.

\begin{figure}
\centering
\begin{tabular}{c}
\includegraphics[width=\columnwidth]{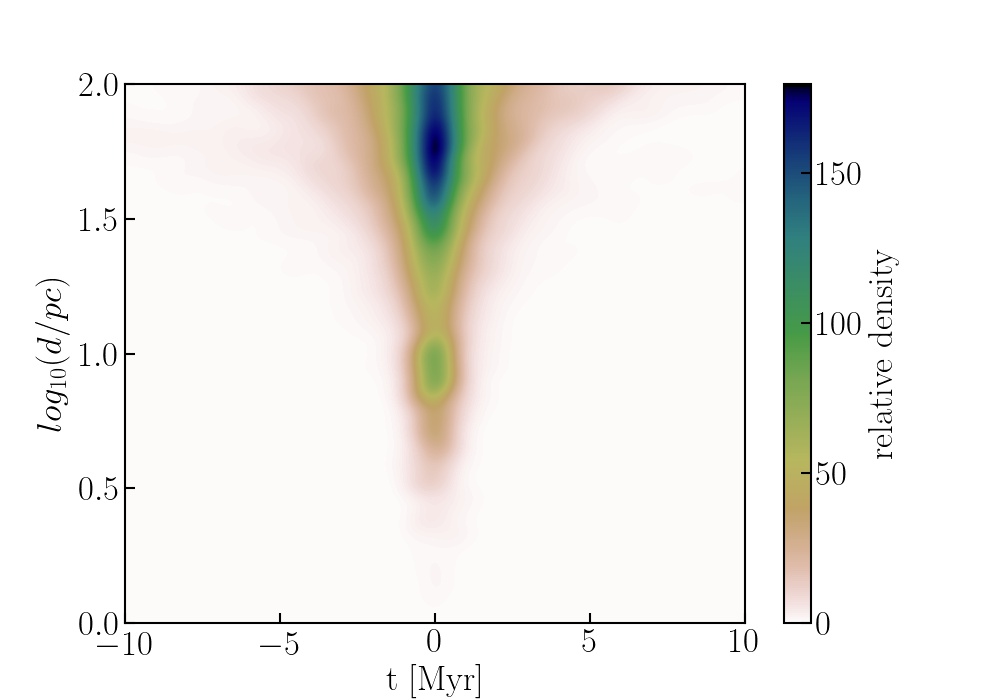}\\
\end{tabular} 
\caption{Distance of closest approach as a function of time since the
current epoch. The color coding shows the relative density. the
color coding is generated using a kernal-density estimator with a
disperison of 1.41\,pc.}
\label{fig:timedist_closest}
\end{figure}

In figure\,\ref{fig:timedist_closest}, we present the distance of the
closest approach to the moment of closest approach. The concentration
to the current epoch (at $t=0$\,Myr) along the x-axis supports the
results presented in figure\,\ref{fig:moment_of_closest_approach} and
\ref{fig:timevel_contours}.  This result is consistent with earlier
estimates of the distance and moment of closest approaching nearby
stars
\citep{2015A&A...575A..35B,2019JPhCS1245a2028D,2020A&A...640A.129W,2020AstL...46..245B}.

When represented in encounter velocity relative to the Sun as a
function of closest-approach distance, in
figure\,\ref{fig:distvel_contours}, the various families are clearly
visible to stay coherent. At the resolution of the simulations,
however, the closest approach distance to a particular asteroid that
originates from a particular star is merely a result of shot-noise due
to our relatively low resolution of 1000 asteroids per \MSun.

\subsubsection{Closest free-floating planets}

\begin{figure}
\centering
\begin{tabular}{c}
\includegraphics[width=\columnwidth]{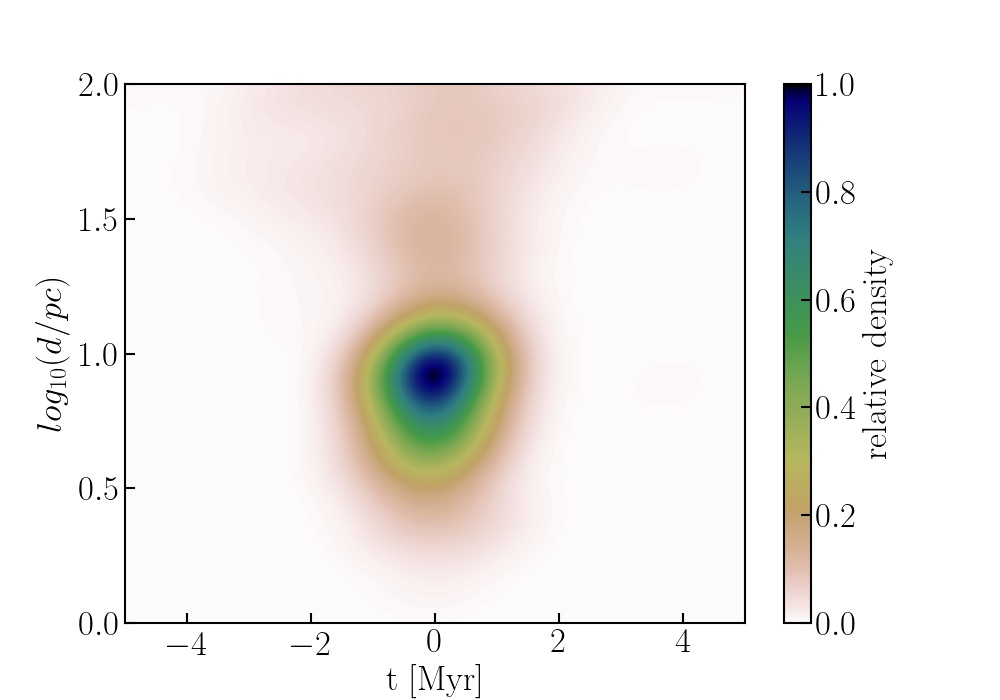}\\
\end{tabular} 
\caption{Distance of closest approach as a function of time since the
current epoch. The color coding shows the relative density. The
color coding is generated using a kernal-density estimator (KDE)
with a disperison of 1.41\,pc. }
\label{fig:timedist_contours}
\end{figure}

In figure\,\ref{fig:timedist_contours}, we present the time of closest
approach and its distance for the planets in our simulations. Whereas
asteroids tend to be spread out over a much large phase-space volume,
planets are confined to the current epoch within about 10\,pc. Massive
planets stay close to their parent stars. Although some are eventually
ejected to become rogue planets, the ejection velocity is so low that
they stick around even more prominently than the asteroids. Asteroids
show a similar over-density, as can be seen in
figure\,\ref{fig:timedist_closest}, but the population is less
pronounced due to the large number of \soli\, at a larger distance and
the small number of asteroids that remain bound to their parent stars.

The relative proximity of free-floating planet to their parent star is
also visible in figure\,\ref{fig:solar_orbit_in_the_galaxy_at_1Gyr},
here planets are more closely surrounding the current Solar position
in the Milky-way Galaxy in comparison to the asteroids.

\section{Discussion}\label{Sect:discussion}

We have simulated a hypothetical population of asteroids and planets
in orbit around the 200 nearest stars to the Sun. In this calculation
we ignored all known planets and binary systems among this population,
but treated each star as single, orbiting in a semi-analytic potential
of the Milkyway Galaxy. Our calculations start 1\,Gyr ago, by
calculating the positions and velocities of the 200 nearby stars
backward with time to this epoch. From there, we surround each star
with 3 or 4 giant planets in circular orbits between 10\,au and
100\,au, and 1000 minor bodies per \MSun\, in the conveyor belt (with
pericenter between 5\,au and 150\,au).

\subsection{Time reversibility and the uniqueness of the calculation}

The integration is continued from 1\,Gyr ago and to 100\,Myr into the
future. This backwards-forwards yo-yo approach is time reversible and
the star, after calculating back to the future, arrives at precisely
the same location with the same velocity as where it is observed
today.

In our simulations, however, the simulated positions and velocities of
individual stars today do not always match the observations by Gaia
DR2, because some of the stars lose one or more planets on the way,
causing its orbit to deviate from the calculated orbit as discussed in
section\,\ref{Sect:IC_B}. The median of the relative difference in a
position perpendicular to the orbit of the star (given the current
solar position this is along the Galactic x-axis) is only $0.39$\,au,
but along the orbit, the discrepancy is $96.7$\,au in the y-direction,
and $451.7$\,au in the galactic z-direction. There are a few cases,
however, for which the ejection of a planet leads to a considerable
deviation of the parent stars' orbit in comparison to the adopted
orbit when calculated without planets. One example is visible in
figure\,\ref{fig:solar_orbit_in_the_galaxy_at_1Gyr} to the left of the
location of today's Solar system. This star, 61~Cyg~A, has an orbit in
the Galactic potential which makes a relatively large angle to the
current Solar motion. Since the dispersion in the orbital calculation
(due to possible ejected planets) is largest along the orbit, this
star is now located rather far from its anticipated location according
to the Gaia data.  Also for other stars, the discrepancy between the
center-of-mass and the actual stellar orbit is largest along the
orbit.

\subsection{The 1\,Gyr time window}\label{Sect:OneGyr}

For the yo-yo approach, we choose a fixed time frame for the
calculations of 1\,Gyr. Ideally, we should have followed each star
since its birth in the Galactic potential to the current time. This
procedure would be hindered by our lack of knowing the ages of all
nearby stars and the long-term evolution of the potential of the Milky
way Galaxy. Another complication arises from the uncertainty in the
actual planet population and the presence of a debris disk.  To
achieve a consistent assessment of the potential contribution to the
observability of the tidal asteroid-tails of these stars would then
require many calculation for each star (to integrate over the
uncertainty in the kinematics and the orientation of the disk) and run
these simulations with a potential model that matches the history of
the Galaxy.

The first analysis would be expensive in terms of computer time, but
not impossible. We simply lack the knowledge of the stellar ages and
do not know the history of the Galaxy well enough to perform such a
simulation.

The tidal tails develop in the first 100\,Myr, and then slowly
diffuse, causing the debris trail to grow in length. This later
diffusion process is slow and develops on a Gyr time scale. For the
length of the tidal-debris tails of the nearest 200 stars, the star's
ages are not critical. This would be different if a particular star is
younger than about 200\,Myr, in which case the star's Oort cloud would
still be in development, and the tidal tail has not yet fully formed.

\subsection{The erosion of the Oort cloud}

According to calculations on the evolution of the Sun's Oort cloud,
\cite{2018MNRAS.473.5432H} derive a half-life of $t_{{1\over 2} OC} =
3$ to $10$\,Gyr.  This range stems from studying figure 4 of
\cite{2018MNRAS.473.5432H} where erosion proceeds faster due to close
encounters with other stars \cite{2011DDA....42.0903D}.

Far-away encounters are not very
important for the erosion of the Oort cloud, but the occasional close
encounter induces a major degradation of the Oort-cloud population.
Without those close encounters, they estimated the Oort cloud half
life to be of the order of $t_{{1\over 2} OC} = 8$ to $10$\,Gyr, which
is an order of magnitude longer than the our estimate of $t_{{1\over
    2} OC} \simeq 800$\,Myr.

Part of the discrepancy can be understood from the lower stellar
masses in our calculations, which measures $\langle m \rangle =
0.72\pm 0.21$\,\MSun. As a result, the Hill radius of the typical
nearby star is smaller than that of the Sun. It will be harder for
these stars to keep asteroids bound. This effect leads to an increase
in the evaporation rate, but cannot explain the order of magnitude
difference.

Another argument for our smaller Oort-cloud lifetime stems from the
orbits of many of the stars, which have somewhat larger ellipticities
than the Sun's orbit around the Galactic center (see
figure\,\ref{fig:Orbit_in_Galaxy_with_OC}). This leads to a strongly
varying Hill radius (see figure\,\ref{fig:RHill_in_Galaxy_with_OC}), a
consequential stronger perturbation by the Galactic tidal field, which
leads to the easier ejection of asteroids \citep{2015A&A...574A..98D}.
We already demonstrated this process to be responsible for the
evaporation of the Oort cloud of Glise 752 in
section\,\ref{Sect:Close_approach}.  On average, this process reduces
the Oort-cloud lifetime but is still insufficient to explain the
discrepancy with \cite{2018MNRAS.473.5432H}.

We argue that the most important difference between the derived
lifetime of the Oort cloud by \cite{2018MNRAS.473.5432H} and our study
stems from how we form the Oort cloud from a population of asteroids
in the conveyor belt, rather than initializing them with a smoothed
and virialized distribution function, as was done in
\cite{2018MNRAS.473.5432H}. The latter method results in a more stable
Oort cloud and a lower evaporation rate.

With our derived half-life on the Oort cloud of $\sim 800$\,Myr, and a
current mass of the Oort cloud of $2$ -- $40$\,M$_{\rm Earth}$
\citep[or $10^{10}$ to $10^{12}$ asteroids, ][]{1996ASPC..107..265W},
we derive an initial mass of the Oort cloud of $3.2$ -- $64$\,M$_{\rm
  Jupiter}$. Which, with an Oort cloud retention fraction of $\sim
0.06$, results in an initial debris disk of $0.003$\,\MSun\, and
0.064\,\MSun. This value is consistent with the initial value of
0.01\,\MSun\, which we adopted in our simulations.  These numbers are
consistent with earlier estimates on the mass of the Oort cloud
\citep{1983A&A...118...90W,1988Sci...242..547M,2004come.book..153D}.

Although ignored in our calculations, occasional close passages of
stars turn out to be important for the Oort cloud's
erosion. Therefore, these passing stars are also expected to be
important for the extent and morphology of the debris streams of
\soli. We ignored this effect in our calculations, and it is hard to
anticipate this effect, except maybe in a statistical sense.

\subsection{Interstellar visitors: 'Omuamua and Borisov}

One of the objectives of this study is to obtain a better
understanding of the contribution of nearby stars to possible
interstellar visitors, such as 1I/'Oumuamua and 2I/Borisov. It was
anticipated that these objects may have originated in the Oort cloud
of other stars \citep{2019AJ....157...86M}.

In an attempt to find the origin of 'Oumuamua and Borisov,
\cite{2018A&A...610L..11D} studied the stars within 60\,pc (using Gaia
DR2 data) but conclude that it would be difficult to find a stellar
origin of the two \soli. The accuracy of the orbits of 'Oumuamua and
Borisov, and the scattering in the Galactic disk make it hard to trace
back their origin beyond some $15$\,pc or within the last $10$\,Myr
\citep{2020AJ....159..147H}.

Despite this earlier analysis, in which nearby stars were excluded as
the source of the two observed \soli\,
\citep{2018MNRAS.479L..17P,2018AJ....156..205B,2020A&A...634A..14B},
\cite{2018A&A...610L..11D} argued that Gl 876 might be a suitable host
of 'Oumuamua. Currently, Glise 876 is at a distance of 4.68\,pc, but
0.817\,Myr ago it had the closest approach to the Sun at a distance of
2.24\,pc. Glise 876 (Gaia DR2 2603090003484152064) regretfully, was
not part of our analysis.  If indeed Glise 876 is a contributor to
interstellar visiting rocks or icy bodies, we might expect multiple
objects to come from the same source.

\begin{figure}
\centering
\hspace*{-0.8cm}
\includegraphics[width=1.0\linewidth]{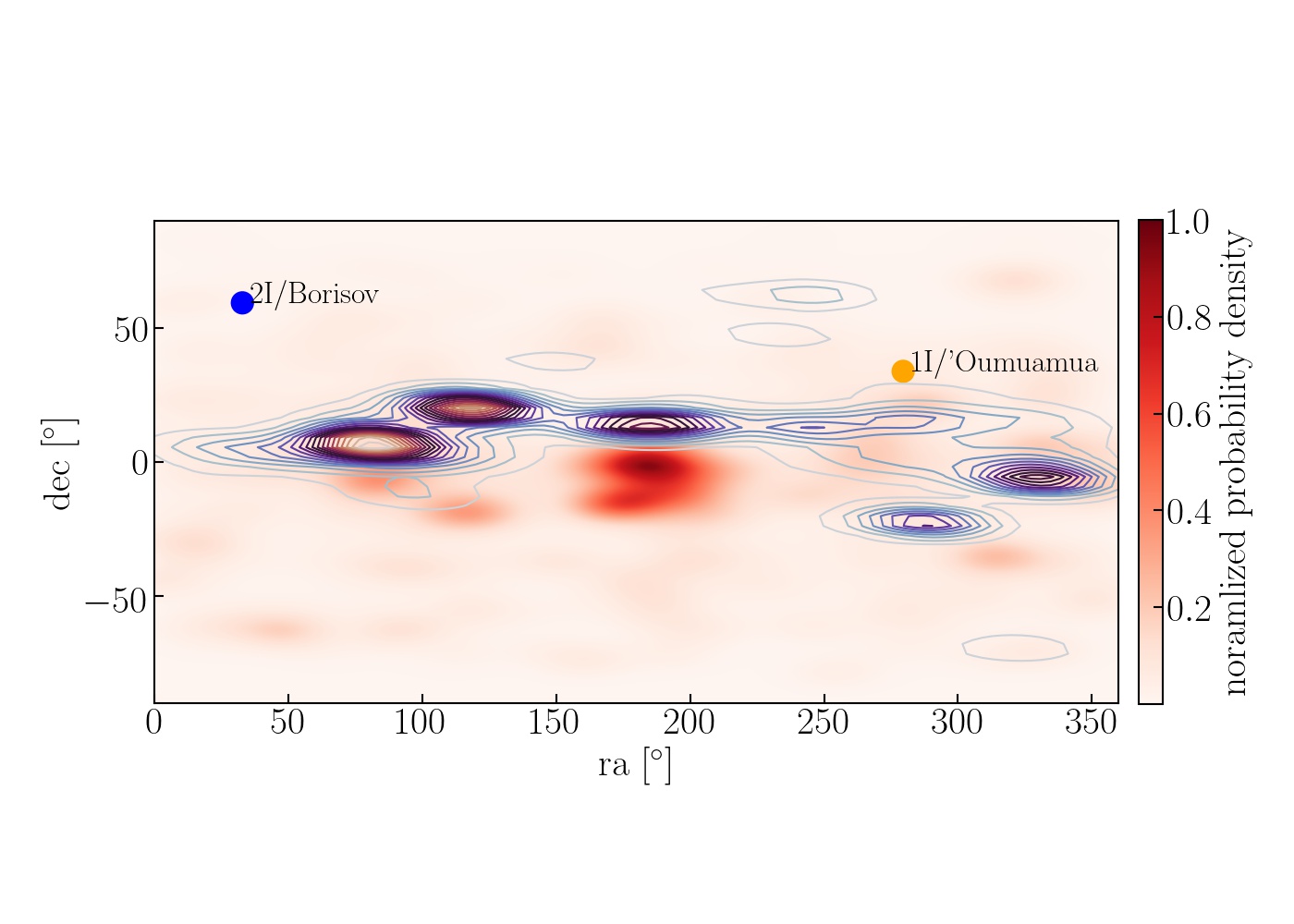}
\caption{Sky projection of nearby asteroids. The blue and orange
bullet points indicates the direction from which Borisov and
'Omuamua entered the Solar system from the analysis of
\cite{2018AJ....156..205B} and \cite{2020A&A...634A..14B},
respectively. The color coding gives the relative contribution to
the local density, wheras the contours give the same information for
the 9\, nearest stars in table\,\ref{tab:nearby_contributors}. }
\label{fig:sky_today}
\end{figure}

Objects such as 'Oumuamua and Borisov, enter the solar system on
unbound orbits and leave again after interacting with the Sun. Both
objects are represented with bullet points in
figure\,\ref{fig:sky_today}, which indicates their point on the sky
from which they appeared to be coming, calculated for 'Oumuamua by
\cite{2018AJ....156..205B} and for Borisov by
\cite{2020A&A...634A..14B}.

Overplotted in figure\,\ref{fig:sky_today} is the distribution of all
minor bodies in our calculations (red shaded region) and of the stars
that contributed the most to nearby objects from
table\,\ref{tab:nearby_contributors} (contours). Borisov is quite far
from these distributions, but 'Oumuamua is approximately near some of
the asteroids that originated from relatively nearby stars.

If multiple objects originate from the same stellar source, they do
not necessarily come from the same direction or with a similar
velocity. If we could argue that multiple intruders have a similar
origin, it may be possible to identify the family and the
star from which they originated. This was illustrated in
figure\,\ref{fig:moment_of_closest_approach}, where we presented
contributing streams of the most contributing stars.

The time 'Oumuamua has been floating around freely in space was
estimated to be less than a Gyr
\citep{2018MNRAS.479L..17P,2019MNRAS.490...21H,2020AJ....159..147H}. In
our calculations, tidal tails remain coherent for considerably
longer. It is then conceivable that 'Oumuamua is still a member of the
leading or trailing arm of debris around its parent star. In that
case, we might expect more objects entering the solar system from the
same Galactic direction.

\begin{figure}
\centering
\hspace*{-0.8cm}
\includegraphics[width=1.0\linewidth]{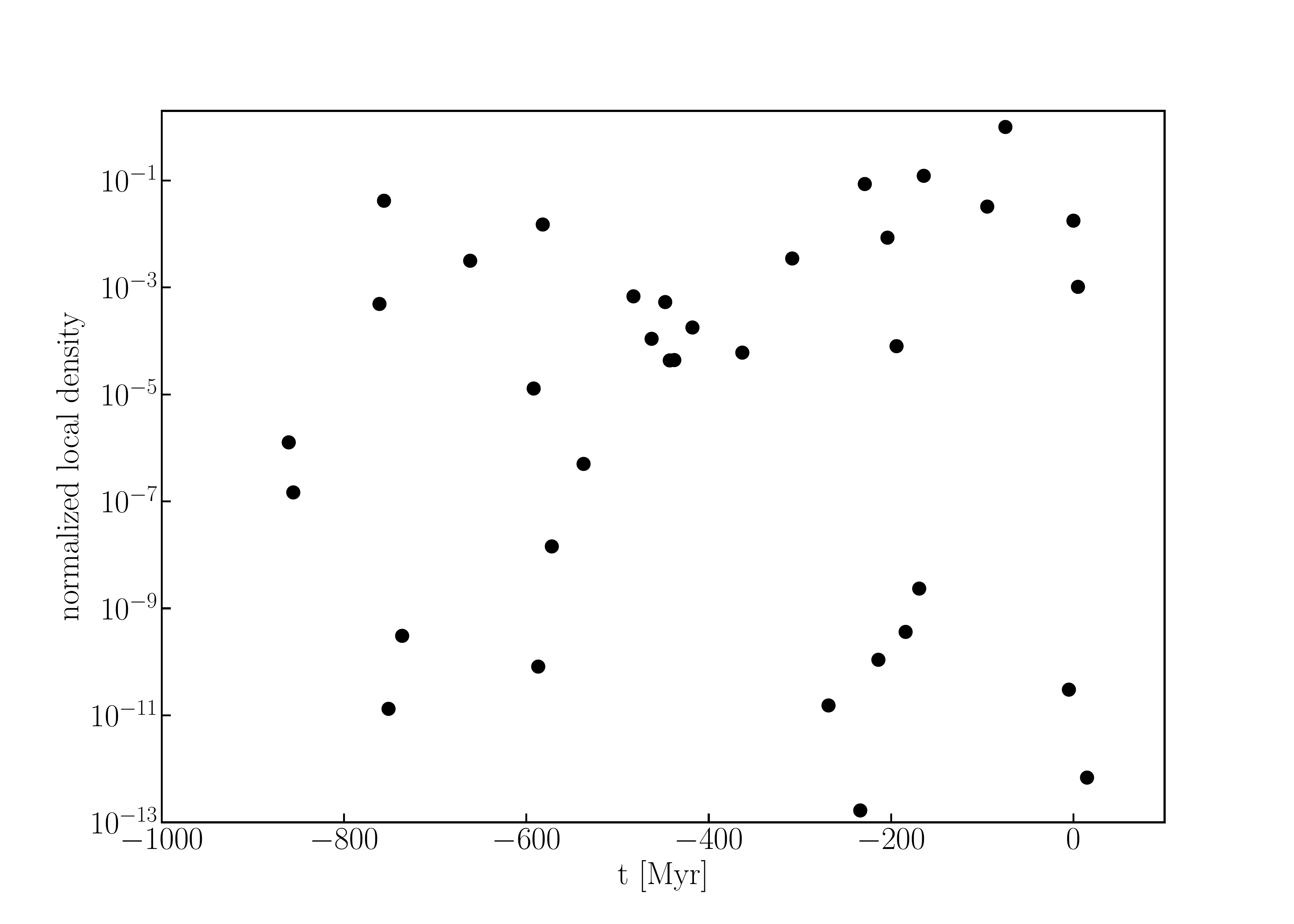}
\caption{Evolution of $\log_{10}$ density of interstellar asteroids in
the Solar vicinity. Calculations were performed with a KDE
smoothing kernel with a kernel bandwidth of 1.4\,pc. Both curves
are normalized to the same maximum of $\sim 0.027$\,M$_{\rm
Earth}$/pc$^{3}$ at today's epoch. The adopted a KDE smooting
kernel of 1\,pc$^{3}$. }
\label{fig:evolution_of_local_density}
\end{figure}

To further verify that the nearby stars contribute prominently, but
that far away stars may still contribute considerably, we present in
figure\,\ref{fig:evolution_of_local_density} the evolution of the
local (and maximum) density of \soli. Today (at $t=0$\,Myr) the peak
reached is about two orders of magnitude higher than at any other
moment. Narrow peaks in time appear when the Solar system
happens to move through the debris-stream of another star.

Today's high peak in the local density of \soli\, does not mean that
we live at a special time, but it is merely a consequence of our
initial selection of the 200 nearby stars.  At any time, there will be
some stars through which's stream we move, giving rise an enhanced
encounter rate with the asteroids in its tidal streams. To what degree
this results in a continuous encounter rate or in rather discrete
peaks is at this point hard to assess.  Based on our simulations, we
cannot determine the temporal variations in this encounter rate
because this depends on the extend and density tructure of the
individual streams. The resolution of our caluclations ware
insufficient to determine this structure, and it would require us to
take many more nearby (and further away) stars into account in the
analysis.

\section{Conclusions}

We have studied the possible evolution of the population of today's
200 nearest stars from the Gaia DR2 catalog in the Galactic potential
including a hypothetical population of initially bound planets and
asteroids. The objective was to characterize the local environment of
the left-overs of the planetary scattering process around these stars
in order to constrain the distribution of asteroids and free-floating
planets in the Sun's vicinity. We ignore the contribution of other
stars, not near to the Sun. Our assumptions on the initial conditions
are somewhat capricious because we do not know if these stars have
debris disks.

For the initial conditions, we assume each star to be born a billion
years ago with 3 or 4 planets in circular orbits between 10 and
100\,au. The planet masses range from 1 to 7 times the mass of
Jupiter. Each planetary system is further populated with 1000
point-mass asteroids per \MSun\, in the plane of the planets. The
orbits have a pericenter distance between 5\,au and 150\,au, and
eccentricities randomly between 0.1 and 0.9. Therefore, they could be
viewed as an excited population that already underwent some scattering
from the planets formation process and early dynamical evolution. Each
planetary system is oriented randomly. The birth location for each
star is determined from back-calculating its orbit in the Galactic
potential.

After setting up the initial conditions, we numerically integrate all
the stars, planets, and asteroids in the Galactic potential for 1.1
billion years, until the current epoch and 100\,Myr into the future.
While performing these calculations, we keep track of the positions of
the Sun and all other objects to find the moment of closest approach
between any object with respect to the Sun.

We find that in the 1 billion years of evolution, 98 percent of the
asteroids become unbound and turn into free-floating asteroids, or
\soli. About 2 percent of the asteroids remain bound to their parent
star in the form of an Oort cloud, and $\sim 0.31$ percent remains
bound to the parent star in stable orbits outside the range of
influence of the outer-most planet, in the parking zone. A tiny
fraction of asteroids, $\sim 0.1$\,\% (164) find themselves bound to
the parent star outside their respective Hill radii.

The fraction of planets that remain bound in the parent star's Oort
cloud is $\sim 2$\,\%, similar to the fraction of asteroids that
remain bound.  This fraction is consistent to the results of the
calculations by \cite{2019MNRAS.490.2495C}.

In the first few tens of millions of years, when the Oort clouds are
filled, the inner and outer Oort clouds have quite distinct
distributions in eccentricity. This changes after about 50\,Myr, when
the eccentricity distribution approaches the thermal distribution, as
demonstrated in figure\,\ref{fig:eccentricity_distribution}.

The distribution in eccentricities of the Oort cloud objects reaches a
steady-state of $P(e) \propto e^{2.75}$ after about 50\,Myr. The
distribution in semi-major axes is flat in $\log$ from $10^4$\,au to
$10^5$\,au.  The half-life of the Oort clouds in our calculations is
800\,Myr, which is considerably shorter than found in earlier
studies. The short Oort cloud half-life in our simulation is
attirbuted to the correlation between semi-major axis, eccentricity,
and inclination.  But some contribution also comes from the lower
masses of the stars and their relatively high eccentricity in the
Galactic potential.

The $10$ stars in the Gaia DR2 database that contribute most
prominently to the probability of producing an interloper are not
necessarily the closest stars. Some relatively distant stars contribute
more prominently than some closer in. This probability depends on the
orbit of the particular star (and its asteroids) with respect to the
Sun. A star that orbits parallel to the Sun in the Galactic potential
is unlikely to contribute to the local density of \soli\, whereas a
crossing orbit results in a local overdensity. A relatively far away
star can then still contribute considerably to the local asteroid
density, in contrast to a more local star. These peaks in over-density
of local \soli\, does not have to coincide with the closest approach
for that particular host star, but it may happen millions of years
earlier or later.

The contribution to the locally observed population of \soli\,
strongly depends on whether or not the Sun happens to move through the
tidal asteroid-tail of another star. Some of these trails are dense
and narrow, and may give rise to a local enhancement in the incident
rate of interstellar objects interacting with the Solar system. We
derive a current local density of \soli\, from the 200 nearest stars
of $\sim 0.027\,M_{\rm Earth}/$pc$^3$. With a typical mass comparable
to 'Oumuamua, we then arrive at a local number-density of $\sim
1.2\times 10^{14}$ per cubic parsec, which is a factor of 10 lower
than earlier estimates based on the two known \soli. We, therefore,
argue that the nearest few hundred stars contribute to the rate of
interstellar asteroids visiting the Solar system, but they are
probably not the dominante source.

We expect considerable structure in the distribution of interstellar
asteroids. If observable, this structure will inform us about the
Galactic potential, the orbits and ages of the nearby stars and the
internal reorganizations their planetary systems might have
endured. On these grounds, we do expect considerable temporal
variations in the encounter rate of \soli.

\subsection{Energy consumption of this calculation}

The evolution of each star (incuding the planets, asteroids and the
Galactic tidal field) was computed in 3 to 7 days on a one GPU and 6
cores. The total computer time spend summs up to about 1000 hours on
GPU and 6000 CPU hours. With 180\,Watt/h per GPU and 12\,Watt/h per
CPU our total energy consumption for the calculations is about
250\,kWh. With $0.283$\,kWh/kg
\citep{DBLP:journals/corr/abs-1304-7664} results in 200 tonnes Co$_2$,
quite comparable to launching a rocket into space
\citep{2020NatAs...4..819P}.

\section*{Public data}
The source code, input files, simulation data and data processing
scripts for this manuscript are available at figshare under DOI {\em
  10.6084/m9.figshare.12834803}.

An animation of the simulation is presented in
\url{https://youtu.be/0fYeAW3e9bQ}.

\section*{Acknowledgment}
It is a pleasure to thank Fransisca Concha-Ram{\'\i}rezs, Santiago
Torres, Anthony Brown, Martijn Wilhelm for discussions.  I thank the
referee for helping considerably in improving the presentation and
shaping the paper.

This work was performed using resources provided by the Academic
Leiden Interdisciplinary Cluster Environment (ALICE).

\end{document}